\documentclass[aps,prl,twocolumn,amsmath,amssymb,superscriptaddress,floatfix,longbibliography]{revtex4-2}
	
\usepackage{amsmath}
\usepackage{txfonts}
\usepackage{microtype}
\usepackage{graphicx}
\usepackage{color}
\usepackage{ulem}
\usepackage{bm}
\usepackage[english]{babel}

\begin{document}

\title{Low-energy photoelectron structures for arbitrary ellipticity of a strong laser field }

\author{Q. Z. Lv}\email{Q.Z.L and M.K. have equal contributions, in numerical, and   analytical calculations, respectively. }
\affiliation{Max-Planck-Institut f\"{u}r Kernphysik, Saupfercheckweg 1,
	69117 Heidelberg, Germany}
	\affiliation{Graduate School, China Academy of Engineering Physics, Beijing 100193, China}
\author{M. Klaiber}\email{Q.Z.L and M.K. have equal contributions, in numerical, and   analytical calculations, respectively. }
\affiliation{Max-Planck-Institut f\"{u}r Kernphysik, Saupfercheckweg 1,
	69117 Heidelberg, Germany}
\author{P.-L. He}
 \affiliation{Max-Planck-Institut f\"{u}r Kernphysik, Saupfercheckweg 1,
	69117 Heidelberg, Germany}
	\affiliation{Key Laboratory for Laser Plasmas (Ministry of Education) and School of Physics and Astronomy, Collaborative Innovation Center for IFSA (CICIFSA), Shanghai Jiao Tong University, Shanghai 200240, China}
 \author{K. Z. Hatsagortsyan}\email{k.hatsagortsyan@mpi-hd.mpg.de}
\author{C. H. Keitel}
\affiliation{Max-Planck-Institut f\"{u}r Kernphysik, Saupfercheckweg 1,
	69117 Heidelberg, Germany}

\date{\today}

\begin{abstract}

Previous attoclock experiments measuring the photoelectron momentum distribution (PMD) via strong-field ionization in an elliptically polarized laser field have shown anomalously large offset angles in the  nonadiabatic regime with large Keldysh parameters ($\gamma$). We investigate the process theoretically in the complete range of ellipticity ($\epsilon$) and large range of $\gamma$, employing numerical solutions of time-dependent Schr\"odinger equation and nonadiabatic classical-trajectory Monte Carlo simulations matched with the under-the-barrier motion via the nonadiabatic strong field approximation. We show the formation of low-energy structures (LES) at any ellipticity value when the Keldysh parameter is sufficiently large. Three regimes of the interaction in the ($\epsilon$-$\gamma$)-space of parameters are identified via the characteristic  PMD features. 
The significant modification of the recollision picture in the nonadiabatic regime,  with so-called anomalous and hybrid slow recollisions, is shown to be behind the LES, inducing extreme nonlinear Coulomb bunching in the phase-space in the polarization plane.
 Our findings elucidate subtle features of the attosecond electron dynamics in strong-field ionization at extreme conditions and indicate limitations on attosecond imaging.

\end{abstract}

\date{\today}

\maketitle

Recollision is a  concept of paramount significance in strong-field physics~\cite{Corkum_1993}. It is behind the fundamental processes of attoscience  \cite{Corkum_2007}, such as high-order harmonic generation (HHG),  nonsequential double-ionization (NSDI), laser-induced electron diffraction, and ultrafast holography \cite{Brabec_2000,Becker_2002,Agostini_2004,Krausz_2009,Meckel_2008,Huismans_2011,Agostini_2024, L'Huillier_2024,Krausz_2024}. Several Coulomb effects are known to be induced by recollisions. Thus, the so-called Coulomb focusing (CF) \cite{Brabec_1996,Comtois_2005,Yudin_2001a} is generated by multiple forward rescattering, which creates a cusp in the transverse photoelectron momentum distribution (PMD) of tunnel-ionized electrons in a linearly polarized laser field \cite{Rudenko_2005}, which shows a counterintuitive shift due to nondipole effects \cite{Ludwig_2014,Danek_2018,Willenberg_2019}. One of the late surprises in strong-field physics  -- low-energy structures (LES) \cite{Blaga_2009,Faisal_2009,Wu_2012b} is also connected with recollisions, namely, with slow forward rescattering in a linearly polarized laser field, and consequent CF and Coulomb longitudinal bunching  \cite{Liu_2010,Yan_2010,Kastner_2012,Lemell_2012,Becker_2014,Wolter_2015x}.

 \begin{figure*}
  \begin{center}
  \includegraphics[width=0.8\textwidth]{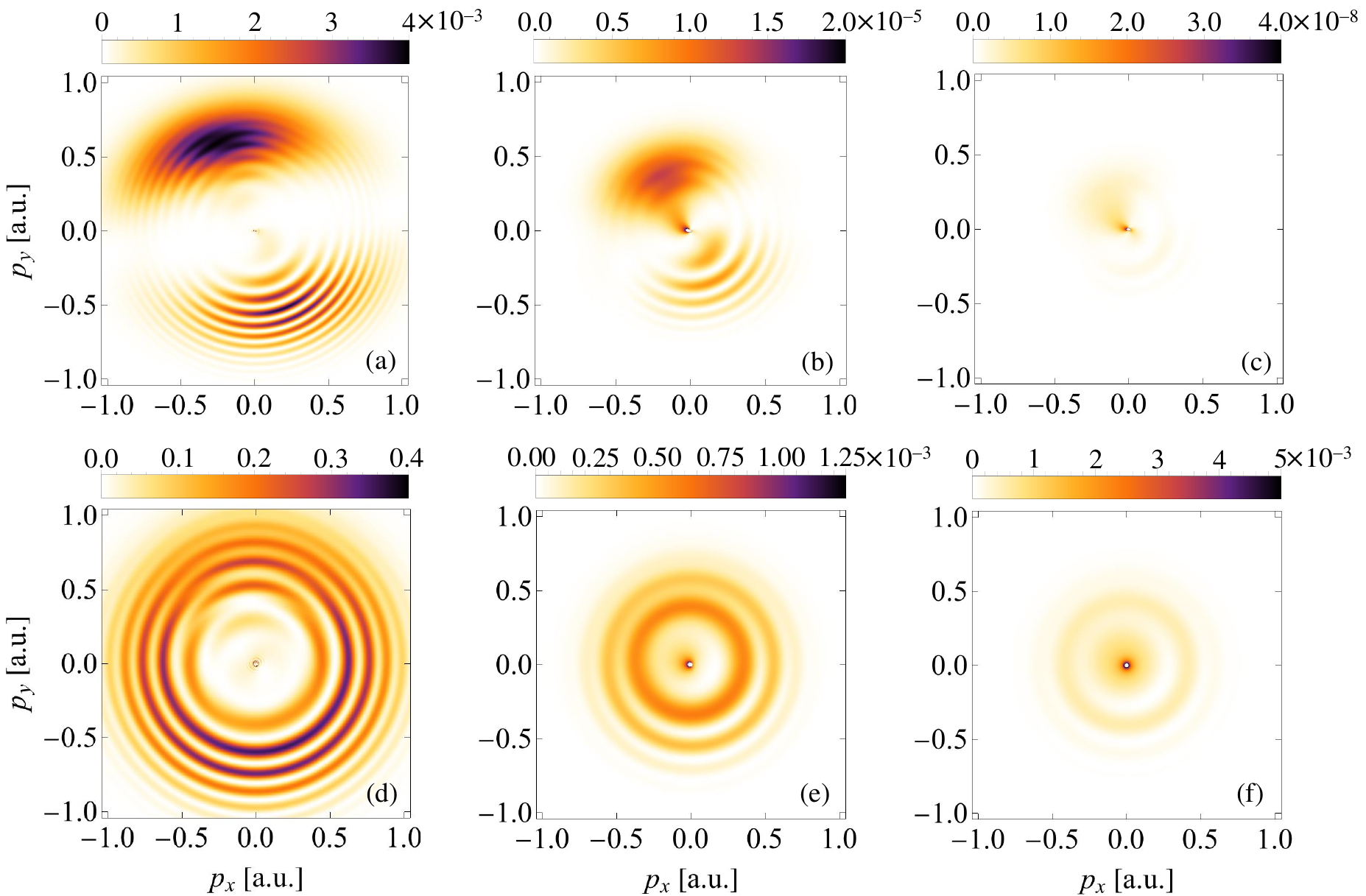}
   \caption{PMDs showing the emergence of LES: First row for $\omega=0.05$, $\epsilon=0.7$: (a) $\gamma=1.2$,  (b) $\gamma= 2$, (c) $\gamma=3.5$. Second  row for $\omega=0.1$, $\epsilon=1$: (a) $\gamma=1.5$,  (b) $\gamma= 3.5$, (c) $\gamma=4$.   Recollisionless (RL) regime, only common elliptical lobes are available [see (a) and (d)]; Recollision Enabled (RE) regime, LES emerges, lobes and LES have peaks of the same order [see (b) and (e)]; Recollision Dominated (RD) regime, only LES is prominent [see (c) and (f)].  Pulse duration is 8 cycles.} 
  \label{PMD}
\end{center}
\end{figure*}

Since early works \cite{Paulus_1998,Kopold_2000,Paulus_2000} it was known that recollisions and, consequently, HHG \cite{Moeller_2012,Lai_2013} and NSDI \cite{Shvetsov-Shilovski_2008}, are possible in an elliptically polarized field with a reduced probability. Wang and Eberly have shown \cite{Wang_2009,Wang_2010} that for the tunneling regime  at the small Keldysh parameters \cite{Keldysh_1965}, $\gamma\ll1$, the elliptical lateral drift can be compensated by the initial transverse momentum of the electron within the tunneled  wave packet, enabling recollisions. Moreover, in the tunneling regime, the recollisions can take place with the same impact parameter and momentum as in a linearly polarized laser field \cite{Maurer_2018}, which yields a cusp in the PMD  similar to the case of linear polarization  \cite{Liu_2012les}, coexisting with PMD elliptical lobes at small ellipticity, and termed as ``sharp edge" in Ref.~\cite{Maurer_2018}.  In particular, the  recollisions at small ellipticities are evidenced in \cite{Shafir_2013,Landsman_2013,Li_2013} by the CF cusp in the lateral momentum distribution for low energies.

An attoclock, based on  strong-field ionization in an elliptically polarized laser field close to circular, has been put forward for time-resolved ionization study with a precision of tens of attoseconds, mapping the tunneling time  delay to the attoclock offset angle \cite{Eckle_2008a,Eckle_2008b,Pfeiffer_2012,Landsman_2014o,Camus_2017,Sainadh_2019}. The large ellipticity is designed to avoid recollision which complicates the mapping. While at relatively small Keldysh parameters ($\gamma \lesssim 1$) the theory reproduces the attoclock offset angle quite accurately \cite{Ivanov_2014,Quan_2019,Serov_2019,Serov_2021}, in the deep nonadiabatic regime at $\gamma \gg 1$, the experiment shows a large offset angle, which is not explained by the common Coulomb momentum transfer at the tunnel exit \cite{Goreslavski_2004,Boge_2013}. The intuitive estimation of the Coulomb effect in the attoclock via the so-called Keldysh-Rutherford model  \cite{Bray_2018} provides a good evaluation for the offset angle at $\gamma \lesssim 1$, but failed in the deep nonadiabatic regime and for small ellipticities \cite{Douguet_2022}. The nonadiabaticity is favorable for recollisions and NSDI in elliptically polarized  fields, as  indicated in many studies \cite{Klaiber_2015,Xie_2024,Fu_2012,Pfeiffer_2012prl,Li_2016,Han_2019,Dubois_2020a,Trabert_2021}. Using the methods of nonlinear dynamics, Uzer \textit{et al.}  predicted  NSDI even in a circularly polarized laser field \cite{Mauger_2010a,Mauger_2010b,Mauger_2013}.   However, a transparent   intuitive picture of the recollision dynamics in the nonadiabatic regime and its role in  LES formation is still missing.  As the  observation of nonadiabatic effects at high  ellipticity  is  hindered by a small ionization yield,  recently, a  paradigm change has been proposed in Ref.~\cite{Heldt_2023},  to initiate the process by an attosecond pump pulse followed by an infrared probe.

In this Letter, we investigate theoretically the recollision dynamics in elliptically polarized laser fields in the full range of ellipticity ($\epsilon$) and a large range of the Keldysh parameter in the nonadiabatic regime. We aim to analyze the emergence of LES at high ellipticity in the nonadiabatic regime, rather than high energy rescattering and NSDI as in Refs.~\cite{Mauger_2010a,Mauger_2010b,Mauger_2013}. The PMDs are calculated employing numerical solutions of the time-dependent Schr\"odinger equation (TDSE). 
For a detailed analysis and intuitive understanding, the nonadiabatic classical-trajectory Monte Carlo (naCTMC) method is developed, where the initial conditions formed during under-the-barrier dynamics are derived via the nonadiabatic strong field approximation (SFA). Three regimes of the interaction in the ($\epsilon$-$\gamma$)-plane of parameters are distinguished and the conditions separating those regions are identified. We show that the recollisions, essentially modified in the nonadiabatic regime, dubbed here as anomalous slow recollisions, and hybrid recollisions, are responsible for the creation of LES  via  Coulomb field-induced bunching in the 2D  phase-space in the laser polarization plane, and the CF in the propagation direction.

We solve numerically TDSE for a hydrogen atom in an elliptically polarized laser field with the vector potential $\mathbf{A}(t)= E_0f(t)(1+\epsilon^2)^{-1/2}\left[\mathbf{e}_x \sin(\omega t)-\epsilon\mathbf{e}_y\cos(\omega t)\right]$, where $\epsilon$ is the ellipticity, $E_0$  the field amplitude, $\omega$ the frequency, and $f(t)$ the pulse shape.
 We consider two laser wavelengths: infrared ($\omega=0.05$) and its second harmonic  ($\omega=0.1$). Atomic units are used throughout. A smooth pulse shape is chosen, see  the Supplemental Materials (SM) \cite{SM}, to avoid edge effects, especially crucial in the nonadiabatic regime $\gamma \gtrsim 1$ \cite{Klaiber_2022edge},

The characteristic PMDs shown in Fig.~\ref{PMD} elucidate three regimes of the interaction:   Recollisionless (RL) regime, with the lobes typical for  attoclock;  Recollision Enabled (RE) regime, with the lobes and LES having peaks of the same order, and  Recollision Dominated (RD) regime with the PMD peak at LES. The regime tends from RL to RD when decreasing $\epsilon$ and/or increasing $\gamma$. In  RE regime,  when decreasing the ellipticity, firstly the tail of LES appears with the peak at the lobe, and then the PMD shows two comparable peaks at the lobe and LES.  At rather small ellipticity in the RD regime, the PMD becomes similar to the case of linear polarization  (e.g. at  $\epsilon <0.3$ for $\gamma=0.3$, $\epsilon <0.4$ for $\gamma=2 $, or at $\epsilon <0.6$ for $\gamma=4$, with $\omega=0.05$), see SM \cite{SM}.

 \begin{figure}[b]
  \begin{center}
\includegraphics[width=0.45\textwidth]{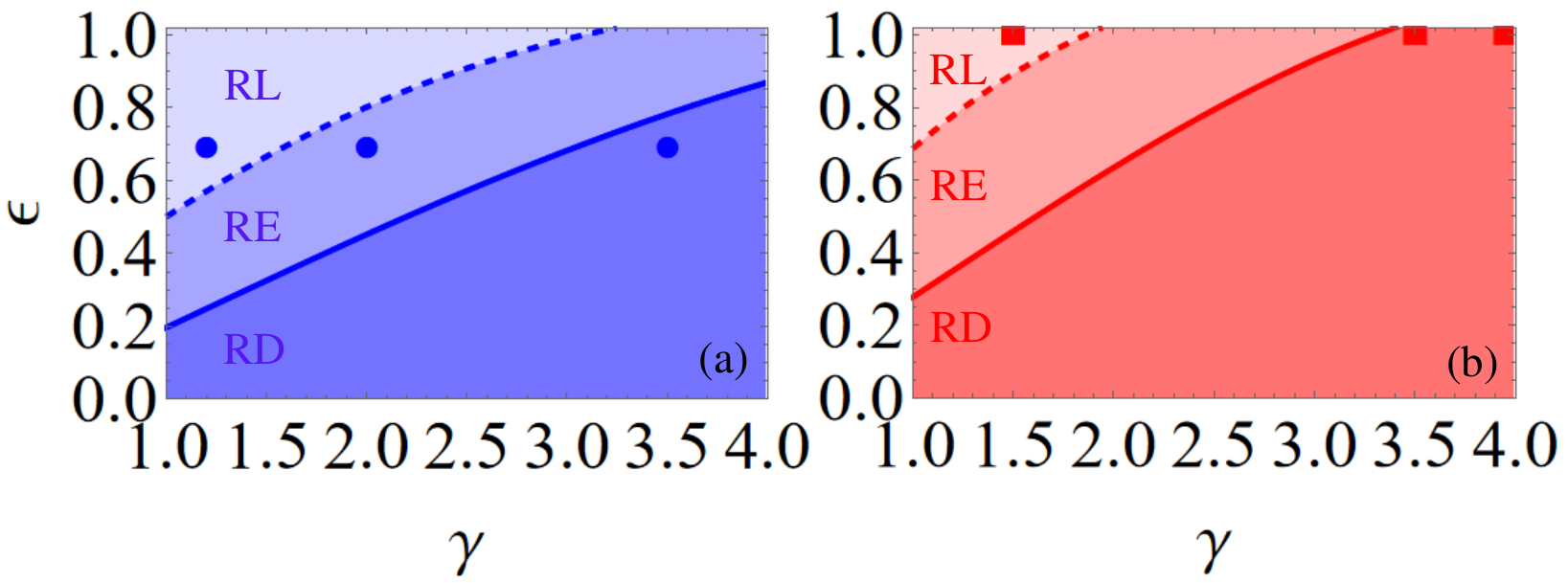}
 \caption{The conditions between the three regimes in ($\epsilon$-$\gamma$)-plane: (dashed) between RL and RE regimes; (solid) between RE and RD regimes,  for $\omega=0.05$ (blue) and $\omega=0.1$ (red). The cycles and squares indicate the PMD parameters of Fig.~\ref{PMD}. }
\label{conditions}
\end{center}
\end{figure}
 \begin{figure}
  \begin{center}
  \includegraphics[width=0.5\textwidth]{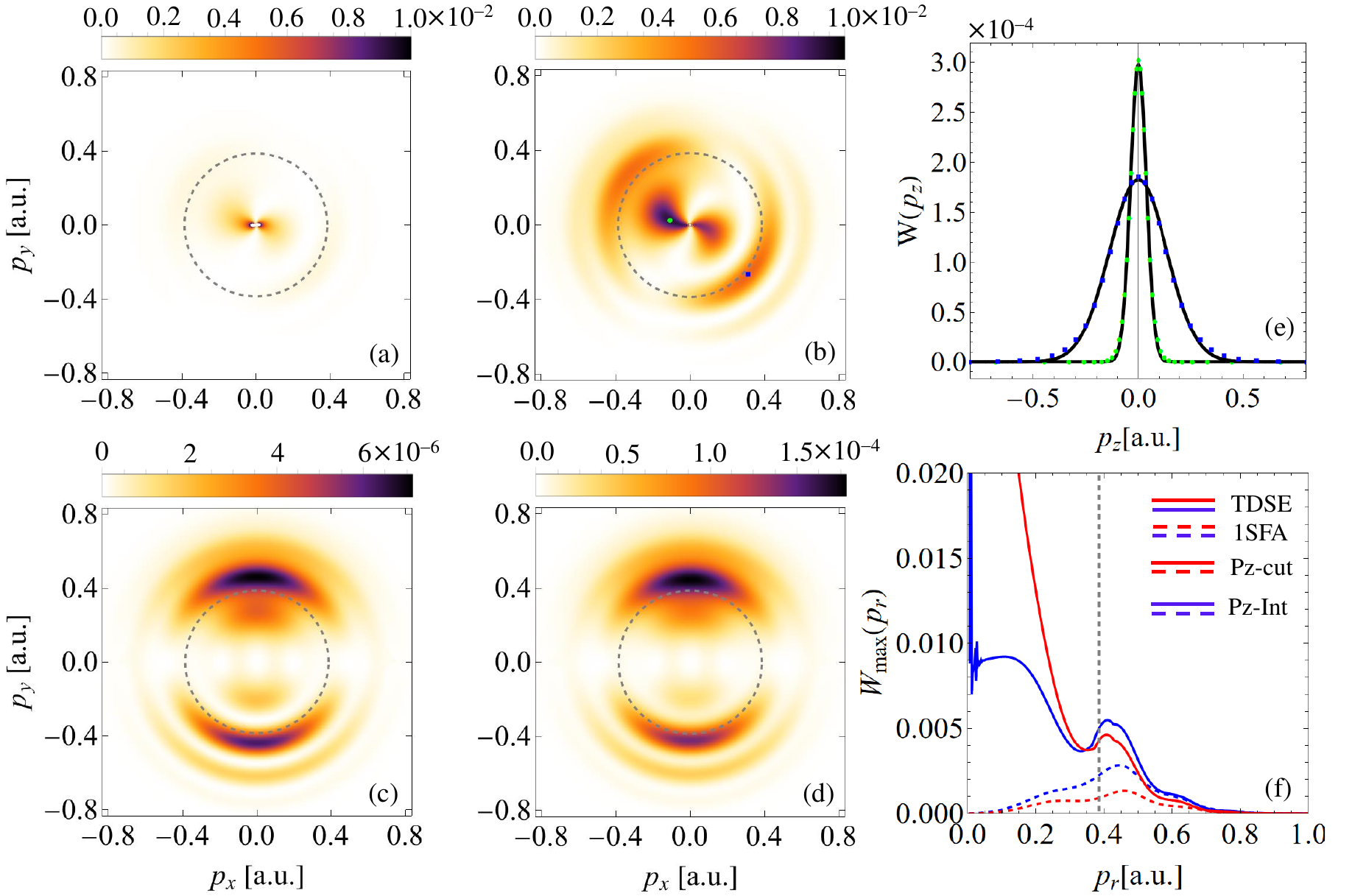}
  \caption{ PMD for $\omega=0.1$, $\gamma=4$,  and $\epsilon=0.8$:  (a)-(b) via TDSE; (c)-(d) via first-order SFA; (a),(c) PMD cut at $p_z=0$; (b),(d) PMD integrated over $p_z$; (e) The  momentum distribution in the propagation direction $p_z$ via TDSE for the PMD point at the LES [green point in (b)] and  at the lobe [(blue point in (b)]; (f) The  maximum probability for each 
  $p_r=\sqrt{p_x^2+p_y^2}$, via TDSE (solid line) and first-order SFA (dashed line) for the  $p_z=0$  cut (red), and  integrated over $p_z$ (blue).  }
  \label{fig3}
\end{center}
\end{figure}
\begin{figure*} 
  \begin{center}
  \includegraphics[width=0.7\textwidth]{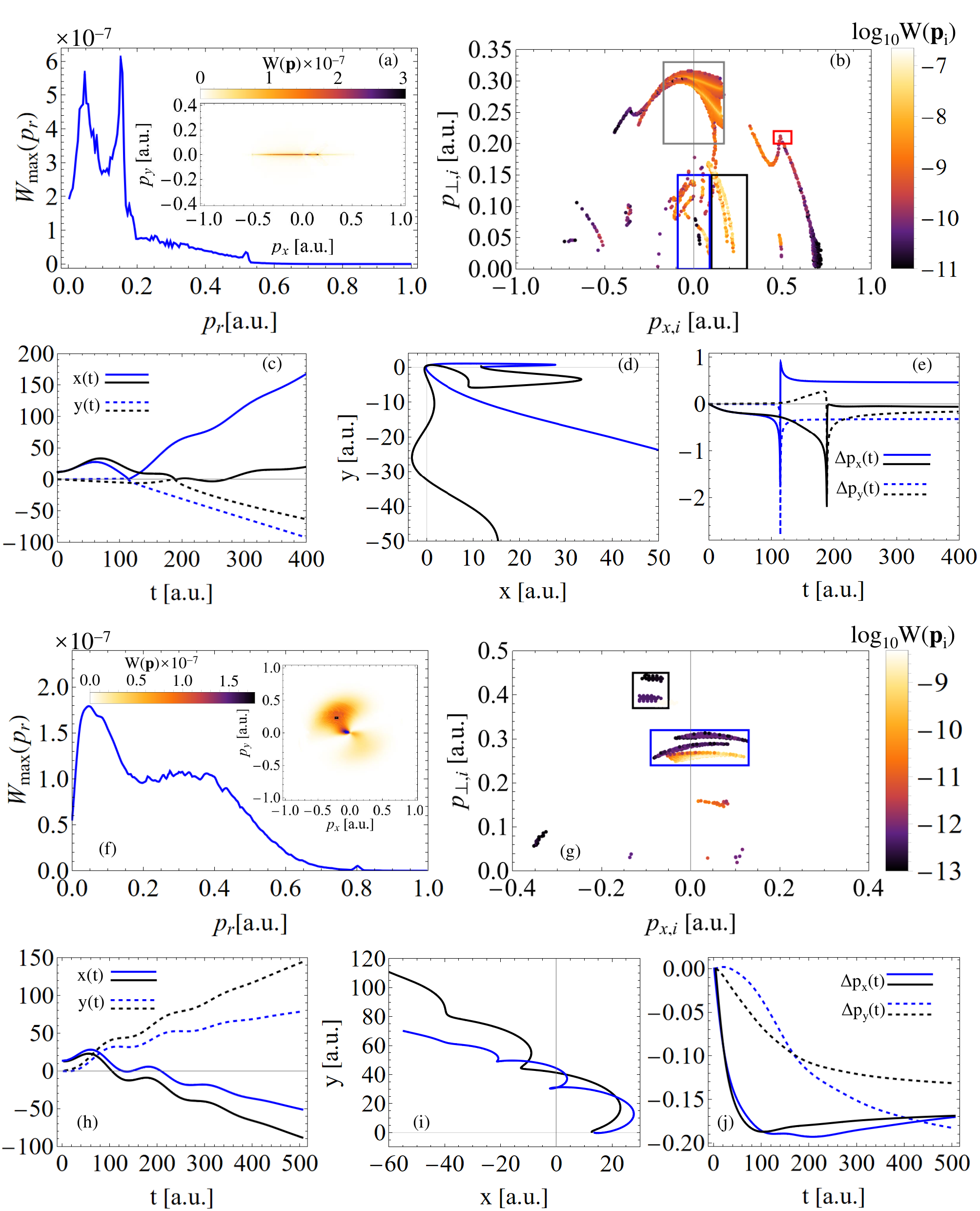}
    \caption{naCTMC simulations in the nonadiabatic regime $\gamma=2$: (a)-(e) for $\epsilon=0$, (f)-(j) for $\epsilon=0.6$. (a) Photoelectron $p_r$ distribution (PrD) (inset shows PMD); (b) The  undisturbed phase space of naCTMC contributing to LES with the color coded probability, blue (black) box mostly contributes to the first (second) peak of PrD,  grey and red boxes contribute to the background;  The features of the typical trajectories: (c) Coordinates, (d) Trajectories, (e) Coulomb momentum transfer, the first (second) anomalous slow recollision, solid (dashed) line. naCTMC results for $\epsilon=0.4$: (f) PrD (inset shows PMD); (g) The  undisturbed phase space contributing to LES, blue (black) contributes to the first (second) PrD peaks, or blue (black) windows in the PMD; (h) Coordinates, (i) Trajectories, (j) Coulomb momentum transfer.   } 
    \label{CTMC}
\end{center}
\end{figure*}

The conditions between regimes in the ($\epsilon$-$\gamma$)-plane of parameters, which depend on the laser frequency, are shown in  Fig.~\ref{conditions}. The derivation of the conditions is outlined below after   the typical contributing trajectories are identified. In the case of $\gamma=2$,  all regimes  of the interaction can exist  with $\omega=0.05$, as well as with $\omega=0.1$, in the corresponding $\epsilon$-regions [Figs.~\ref{PMD} and \ref{conditions}]. However, for $\omega=0.1$ and $\gamma=4$, the RD regime is always the case at any $\epsilon$.

The nonadiabatic LES at high ellipticity is especially  evident when comparing the TDSE results with that of the first-order SFA, see Fig.~\ref{fig3}. While LES is absent in the SFA calculations [Fig.~\ref{fig3}(c,d)], the LES near zero energy is visible as in the $p_z=0$ cut of PMD [Fig.~\ref{fig3}(a)], as well as in $p_z$ integrated spectrum [Fig.~\ref{fig3}(b)], and in the radial $p_r=\sqrt{p_x^2+p_y^2}$ momentum distribution [Fig.~\ref{fig3}(f)], which features a large LES peak at $p_r=0$. The narrow  momentum distribution along the laser propagation direction  for LES electrons compared to the lobe ones [Fig.~\ref{fig3}(e)] is indicative of CF in the lateral direction \cite{Shafir_2013}.

For an intuitive explanation of elliptical LES, we invoke naCTMC. The analysis of the typical trajectories creating LES in the nonadiabatic regime $\gamma>1$ (in contrast to the tunneling regime  $\gamma \ll 1$), we begin with the case of linear polarization $\epsilon=0$ at $\gamma=2$. In Fig.~\ref{CTMC}(a) we show the PMD and $p_r$ distributions (PrD) which demonstrate two prominent LES peaks. While LES peaks in the adiabatic regime arise during the normal slow recollisions [with vanishing longitudinal velocity at the recollision of the \textit{long trajectories}: the coordinate $x_e$ of the tunnel exit and the asymptotic momentum are of opposite sign] each LES peak corresponding to the first, second,\textit{ etc.} slow recollisions when decreasing the final energy \cite{Liu_2010,Kastner_2012}, here in the nonadiabatic regime the picture is different. We  analyze this in Fig.~\ref{CTMC}(b) via the probability distribution over the undisturbed phase space of naCTMC (the final PMD neglecting the Coulomb field in the simulation)  for $\gamma=2$.

There are the following main differences from the case of LES at $\gamma\ll 1$ \cite{Liu_2010}. While the inner phase space contribution to the nonadiabatic LES ($\gamma\gg 1$) [$p_{\bot ,i}=\sqrt{p_{y,i}^2+p_{z,i}^2} <0.2$ in Fig.~\ref{CTMC}(b), blue and black boxes] dominates over that of the outer phase spaces [$p_{\bot ,i} >0.2$, grey box], in the common LES ($\gamma\ll 1$), only the outer phase space has significant contribution to the PMD. The phase space in the blue box is mostly responsible for creating the first low energy peak in PrD in Fig.~\ref{CTMC}(a), while the black box phase space is for the PrD second peak. Furthermore, the recolliding trajectories contributing to LES are modified.  Specific anomalous slow recollisions arise in the nonadiabatic regime, responsible for the LES peaks in Fig.~\ref{CTMC}(a). The dominant trajectories from the blue (black) boxes are shown in Fig.~\ref{CTMC}(c). They are typical anomalous slow recollisions [with vanishing longitudinal velocity at the recollision of \textit{short trajectories}: the coordinate $x_e$  and the asymptotic momentum are of the same sign, see Fig.~\ref{CTMC}(c)], the blue trajectory with first slow recollision, and the black with second slow recollision [Fig.~\ref{CTMC}(d,e)] contributing to the first and the second PrD peaks, respectively. This kind of anomalous slow recollisions are contributing at nonvanishing final momenta only at large $\gamma\gtrsim 2$ with the exit coordinate $x_e\sim E_0/\omega^2$. The trajectories from the gray and red boxes create bakground of the PrD distribution and their specificity in the nonadiabatic regime are discussed in SM \cite{SM}.

\begin{figure*}  
\begin{center}
\includegraphics[width=1\textwidth]{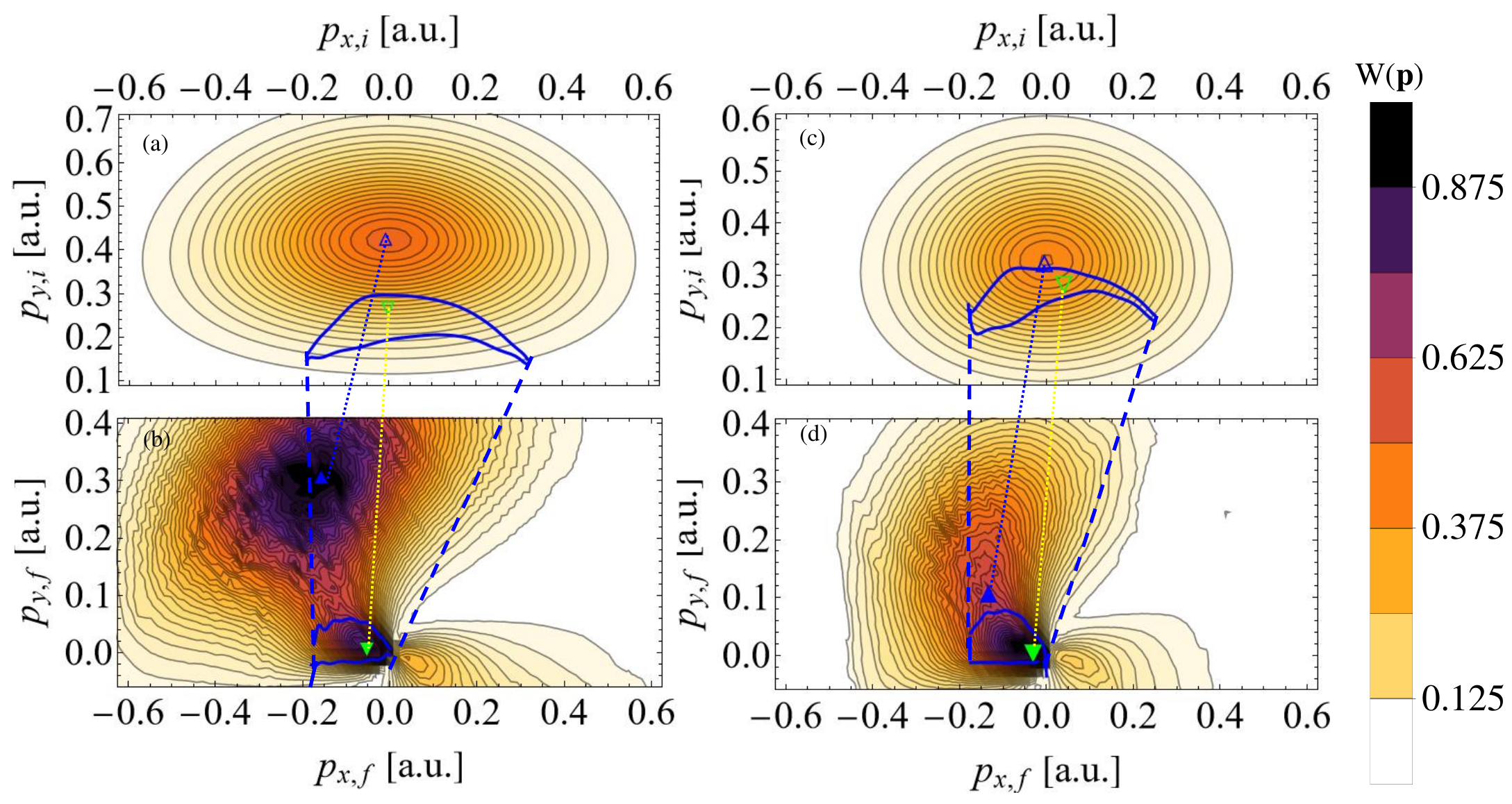}
\caption{The formation of LES in the case of $\omega=0.05$  and $\epsilon=0.6$: (a,b) RE regime,  with $\gamma=2$; (c,d) RD regime with $\gamma=3.2$. The phase spaces in the polarization plane $(p_{x,i},p_{y,i}) $ undisturbed by the Coulomb field  are shown in (a,c), which correspond to PMDs  presented in (b,d). The LES in (b,d)  and the corresponding initial phase spaces in (a,c) are indicated by blue lines. The peak of  the final distribution (green, inverted triangle) does not correspond to the trajectory with the maximum probability in the initial phase space (blue triangle).  }
      \label{2D-bunching}
\end{center}
\end{figure*}

When increasing ellipticity, see the case $\epsilon=0.6$ in Fig.~\ref{CTMC}(f-j)], the role of the inner initial phase space and the anomalous slow recollisions  gradually decrease [Fig.~\ref{CTMC}(g)]. Moreover, the recollisions at large ellipticity in the nonadiabatic regime acquire a hybrid character. While in the common recollision ($\gamma\ll 1$) the Coulomb momentum transfer (CMT) takes place only during the brief time of the rescattering, in the hybrid recollision (at $\gamma\gg 1$)  the CMT is monotonously built-up during the motion from the tunnel exit up to the recollision [Fig.~\ref{CTMC}(j)]. The low (high) energy peak in PrD [Fig.~\ref{CTMC}(f)] is created by the blue (black) boxes in phase space [Fig.~\ref{CTMC}(g)]. The low energy peak in PrD is the LES due to the Coulomb effect, while the second shoulder at high energies is similar to the common attoclock lobe. The typical trajectories feature hybrid recollisions: the blue with mostly two recollisions and a large CMT, the black with mostly single recollision  and smaller CMT [Fig.~\ref{CTMC}(h-j)]. 
Thus, the typical trajectory contributing to the elliptical LES [Fig.~\ref{CTMC}(i)] has a duration of the laser period.

With the given information on the typical recollision trajectory, we have estimated the conditions between the regimes [Fig.~\ref{conditions}] as follows. The electron ionized at the peak of the elliptically polarized laser field will end up with a large momentum $p_{y}^{(\epsilon)}=\epsilon E_0/\omega$ due to the elliptical drift, if there are no recollisions and Coulomb effects, and if the electron is ionized with a vanishing  momentum. The recollisions and LES will be possible if the elliptical drift can be compensated. In the adiabatic tunneling regime ($\gamma\ll 1$) it is compensated by the initial transverse momentum within the transverse momentum width $\Delta^{(a)}_\bot=\sqrt{E_0/\kappa}$ of the Perelomov-Popov-Terent'ev (PPT) adiabatic distribution \cite{Popov_2004u}, with $\kappa=\sqrt{2I_p}$, and the ionization potential $I_p$, yielding the condition  $p_{y}^{(\epsilon)} \lesssim \Delta^{(a)}_\bot$  \cite{Maurer_2018}. In the nonadiabatic regime,  the  initial conditions at the tunnel exit are modified: the peak of the initial transverse momentum is shifted $p_{\bot ,e}=\epsilon \gamma\kappa/6$  \cite{Klaiber_2015}, and the transverse momentum width is increased with respect to the adiabatic case ~\cite{Mur_2001}, e.g. $\Delta_\bot^{(na)}\sim \Delta^{(a)}_\bot\sqrt{\frac{\gamma}{\ln(\gamma+\sqrt{1+\gamma^2})}}$, at  $\epsilon \ll 1$. In our naCTMC they are derived explicitly from the nonadiabatic  SFA \cite{Klaiber_2013a}. The first condition between the RL and RE regimes corresponds to the case when the large initial transverse momentum  within the  momentum distribution at the tunnel exit $\Delta_\bot-p_{y,e}$, and the Coulomb momentum transfer at the recollision ($\delta p_y^C$) counteract the elliptical drift, yielding focusing to the final small momentum associated with LES:
\begin{eqnarray}
p_{y, f}=\epsilon E_0/\omega+p_{y,e} -\Delta_\bot^{(na)}-\delta p_y^C\approx 0.\label{C1}
\end{eqnarray}
The second condition between the RE and RD regimes corresponds to the case when  the electron ionized at the peak of the transverse wave packet ($ p_{y0}$) will be able to reach the LES region:
\begin{eqnarray}
p_{y, f}=\epsilon E_0/\omega+p_{y,e}-\delta p_y^C\approx 0.\label{C2}
\end{eqnarray}
In deriving the explicit conditions via Eqs.~(\ref{C1})-(\ref{C2}), we calculate $\Delta_\bot^{(na)}$ and $p_{y,e}$ via the nonadiabatic Coulomb-corrected SFA. Note that the Coulomb effect $\delta p_y^C$ depends on the initial momentum. Taking into account  that the recolliding trajectory is ionized near the peak of the laser field and the recollision happens after one period of the motion,  the Coulomb momentum transfer is estimated as
\begin{eqnarray}
\delta p_y^C\approx \int dt \frac{Z y_r}{\sqrt{x(t)^2+y_r^2}},
\label{CC}
\end{eqnarray}
with the charge $Z$ of the atomic core, the transverse coordinate at the recollision  $y_r\approx p_{y,r}/\omega$, and $x(t)\approx (E_0/\omega) t$. During LES formation $p_{y,r}\approx \delta p_y^C$, because after the recollision, and the Coulomb momentum transfer, the electron should appear with vanishing momentum. Thus, the variable $\delta p_y^C$ is derived solving Eq.~(\ref{CC}) with respect to it. The Eqs.~(\ref{C1}) and (\ref{C2}) lead to the conditions presented in Fig.~\ref{conditions}.

Finally, we illustrate in Fig.~\ref{2D-bunching} the intuitive picture of LES creation due to the Coulomb effects. While the Coulomb focusing in the laser propagation direction $z$ is physically similar to the adiabatic case, the 2D momentum bunching in the polarization plane  in the case of elliptical polarization in the nonadiabatic regime is more nontrivial and elucidated in Fig.~\ref{2D-bunching}. When commonly discussing the attoclock offset angle, one assumes that the trajectory of the ionized electron with the maximal probability at the tunnel exit will end up with an asymptotic momentum shifted by the corresponding CMT, which will create the  peak in PMD. However, Fig.~\ref{2D-bunching} demonstrates that additionally the phase space transformation due to the Coulomb effect, described by the Jacobian $\partial(p_{x,i},p_{y,i}) /\partial(p_x, p_y)$, is essentially modifying the probability distribution. As a result, the peak of the phase space distribution in the final distribution in the polarization plane does not corresponds to the trajectory with the maximum probability in the phase space undisturbed by Coulomb field, see also SM \cite{SM}.  
As the Coulomb effects significantly distort the phase space in the RE and RD regimes, which essentially changes the position of the PMD peak, the attoclock offset angle can be introduced meaningfully only for the RL regime.

In conclusion, we have investigated the recollision picture in an elliptically polarized laser field in the nonadiabatic regime and identified the specific recolliding trajectories, i.e. the anomalous slow recollisions and the hybrid recollisions which are behind the LES in this regime. We have shown that the LES induced by such recollisions arises at any ellipticity of the laser field if the Keldysh parameter of the nonadiabatic regime is sufficiently large. Three regimes in the $(\epsilon, \gamma)$ plane are classified with different roles of the Coulomb effects. Our findings are important for accurately distinguishing the true tunneling time delay from the Coulomb-induced effect in strong field ionization in different setups \cite{Klaiber_2022R,Yu_2022,Xie_2024}. Our results indicate that at extreme conditions the LES could complicate retrieval of interference fringes and encoded structural information in strong-field elliptical holography in the nonadiabatic regime.

\bibliography{strong_fields_bibliography}

\begin{thebibliography}{74}%
\makeatletter
\providecommand \@ifxundefined [1]{%
 \@ifx{#1\undefined}
}%
\providecommand \@ifnum [1]{%
 \ifnum #1\expandafter \@firstoftwo
 \else \expandafter \@secondoftwo
 \fi
}%
\providecommand \@ifx [1]{%
 \ifx #1\expandafter \@firstoftwo
 \else \expandafter \@secondoftwo
 \fi
}%
\providecommand \natexlab [1]{#1}%
\providecommand \enquote  [1]{``#1''}%
\providecommand \bibnamefont  [1]{#1}%
\providecommand \bibfnamefont [1]{#1}%
\providecommand \citenamefont [1]{#1}%
\providecommand \href@noop [0]{\@secondoftwo}%
\providecommand \href [0]{\begingroup \@sanitize@url \@href}%
\providecommand \@href[1]{\@@startlink{#1}\@@href}%
\providecommand \@@href[1]{\endgroup#1\@@endlink}%
\providecommand \@sanitize@url [0]{\catcode `\\12\catcode `\$12\catcode
  `\&12\catcode `\#12\catcode `\^12\catcode `\_12\catcode `\%12\relax}%
\providecommand \@@startlink[1]{}%
\providecommand \@@endlink[0]{}%
\providecommand \url  [0]{\begingroup\@sanitize@url \@url }%
\providecommand \@url [1]{\endgroup\@href {#1}{\urlprefix }}%
\providecommand \urlprefix  [0]{URL }%
\providecommand \Eprint [0]{\href }%
\providecommand \doibase [0]{https://doi.org/}%
\providecommand \selectlanguage [0]{\@gobble}%
\providecommand \bibinfo  [0]{\@secondoftwo}%
\providecommand \bibfield  [0]{\@secondoftwo}%
\providecommand \translation [1]{[#1]}%
\providecommand \BibitemOpen [0]{}%
\providecommand \bibitemStop [0]{}%
\providecommand \bibitemNoStop [0]{.\EOS\space}%
\providecommand \EOS [0]{\spacefactor3000\relax}%
\providecommand \BibitemShut  [1]{\csname bibitem#1\endcsname}%
\let\auto@bib@innerbib\@empty
\bibitem [{\citenamefont {Corkum}(1993)}]{Corkum_1993}%
  \BibitemOpen
  \bibfield  {author} {\bibinfo {author} {\bibfnamefont {P.~B.}\ \bibnamefont
  {Corkum}},\ }\bibfield  {title} {\bibinfo {title} {Plasma perspective on
  strong field multiphoton ionization},\ }\href@noop {} {\bibfield  {journal}
  {\bibinfo  {journal} {Phys. Rev. Lett.}\ }\textbf {\bibinfo {volume} {71}},\
  \bibinfo {pages} {1994} (\bibinfo {year} {1993})}\BibitemShut {NoStop}%
\bibitem [{\citenamefont {Corkum}\ and\ \citenamefont
  {Krausz}(2007)}]{Corkum_2007}%
  \BibitemOpen
  \bibfield  {author} {\bibinfo {author} {\bibfnamefont {P.~B.}\ \bibnamefont
  {Corkum}}\ and\ \bibinfo {author} {\bibfnamefont {F.}~\bibnamefont
  {Krausz}},\ }\bibfield  {title} {\bibinfo {title} {Attosecond science},\
  }\href@noop {} {\bibfield  {journal} {\bibinfo  {journal} {Nature Phys.}\
  }\textbf {\bibinfo {volume} {3}},\ \bibinfo {pages} {381} (\bibinfo {year}
  {2007})}\BibitemShut {NoStop}%
\bibitem [{\citenamefont {Brabec}\ and\ \citenamefont
  {Krausz}(2000)}]{Brabec_2000}%
  \BibitemOpen
  \bibfield  {author} {\bibinfo {author} {\bibfnamefont {T.}~\bibnamefont
  {Brabec}}\ and\ \bibinfo {author} {\bibfnamefont {F.}~\bibnamefont
  {Krausz}},\ }\bibfield  {title} {\bibinfo {title} {Intense few-cycle laser
  fields: Frontiers of nonlinear optics},\ }\href
  {https://doi.org/10.1103/RevModPhys.72.545} {\bibfield  {journal} {\bibinfo
  {journal} {Rev. Mod. Phys.}\ }\textbf {\bibinfo {volume} {72}},\ \bibinfo
  {pages} {545} (\bibinfo {year} {2000})}\BibitemShut {NoStop}%
\bibitem [{\citenamefont {Becker}\ \emph {et~al.}(2002)\citenamefont {Becker},
  \citenamefont {Grasbon}, \citenamefont {Kopold}, \citenamefont
  {Milo\u{s}evi\'c}, \citenamefont {Paulus},\ and\ \citenamefont
  {Walther}}]{Becker_2002}%
  \BibitemOpen
  \bibfield  {author} {\bibinfo {author} {\bibfnamefont {W.}~\bibnamefont
  {Becker}}, \bibinfo {author} {\bibfnamefont {F.}~\bibnamefont {Grasbon}},
  \bibinfo {author} {\bibfnamefont {R.}~\bibnamefont {Kopold}}, \bibinfo
  {author} {\bibfnamefont {D.~B.}\ \bibnamefont {Milo\u{s}evi\'c}}, \bibinfo
  {author} {\bibfnamefont {G.~G.}\ \bibnamefont {Paulus}},\ and\ \bibinfo
  {author} {\bibfnamefont {H.}~\bibnamefont {Walther}},\ }\bibfield  {title}
  {\bibinfo {title} {Above-threshold ionization: from classical features to
  quantum effects},\ }\href@noop {} {\bibfield  {journal} {\bibinfo  {journal}
  {Adv. Atom. Mol. Opt. Phys.}\ }\textbf {\bibinfo {volume} {48}},\ \bibinfo
  {pages} {35} (\bibinfo {year} {2002})}\BibitemShut {NoStop}%
\bibitem [{\citenamefont {Agostini}\ and\ \citenamefont
  {DiMauro}(2004)}]{Agostini_2004}%
  \BibitemOpen
  \bibfield  {author} {\bibinfo {author} {\bibfnamefont {P.}~\bibnamefont
  {Agostini}}\ and\ \bibinfo {author} {\bibfnamefont {L.~F.}\ \bibnamefont
  {DiMauro}},\ }\bibfield  {title} {\bibinfo {title} {The physics of attosecond
  light pulses},\ }\href@noop {} {\bibfield  {journal} {\bibinfo  {journal}
  {Rep. Prog. Phys.}\ }\textbf {\bibinfo {volume} {67}},\ \bibinfo {pages}
  {813} (\bibinfo {year} {2004})}\BibitemShut {NoStop}%
\bibitem [{\citenamefont {Krausz}\ and\ \citenamefont
  {Ivanov}(2009)}]{Krausz_2009}%
  \BibitemOpen
  \bibfield  {author} {\bibinfo {author} {\bibfnamefont {F.}~\bibnamefont
  {Krausz}}\ and\ \bibinfo {author} {\bibfnamefont {M.}~\bibnamefont
  {Ivanov}},\ }\bibfield  {title} {\bibinfo {title} {Attosecond physics},\
  }\href@noop {} {\bibfield  {journal} {\bibinfo  {journal} {Rev. Mod. Phys.}\
  }\textbf {\bibinfo {volume} {81}},\ \bibinfo {pages} {163} (\bibinfo {year}
  {2009})}\BibitemShut {NoStop}%
\bibitem [{\citenamefont {Meckel}\ \emph {et~al.}(2008)\citenamefont {Meckel},
  \citenamefont {Comtois}, \citenamefont {Zeidler}, \citenamefont {Staudte},
  \citenamefont {Pavicic}, \citenamefont {Bandulet}, \citenamefont {Pepin},
  \citenamefont {Kieffer}, \citenamefont {Dorner}, \citenamefont {Villeneuve},\
  and\ \citenamefont {Corkum}}]{Meckel_2008}%
  \BibitemOpen
  \bibfield  {author} {\bibinfo {author} {\bibfnamefont {M.}~\bibnamefont
  {Meckel}}, \bibinfo {author} {\bibfnamefont {D.}~\bibnamefont {Comtois}},
  \bibinfo {author} {\bibfnamefont {D.}~\bibnamefont {Zeidler}}, \bibinfo
  {author} {\bibfnamefont {A.}~\bibnamefont {Staudte}}, \bibinfo {author}
  {\bibfnamefont {D.}~\bibnamefont {Pavicic}}, \bibinfo {author} {\bibfnamefont
  {H.~C.}\ \bibnamefont {Bandulet}}, \bibinfo {author} {\bibfnamefont
  {H.}~\bibnamefont {Pepin}}, \bibinfo {author} {\bibfnamefont {J.~C.}\
  \bibnamefont {Kieffer}}, \bibinfo {author} {\bibfnamefont {R.}~\bibnamefont
  {Dorner}}, \bibinfo {author} {\bibfnamefont {D.~M.}\ \bibnamefont
  {Villeneuve}},\ and\ \bibinfo {author} {\bibfnamefont {P.~B.}\ \bibnamefont
  {Corkum}},\ }\bibfield  {title} {\bibinfo {title} {{Laser-Induced Electron
  Tunneling and Diffraction}},\ }\href@noop {} {\bibfield  {journal} {\bibinfo
  {journal} {Science}\ }\textbf {\bibinfo {volume} {320}},\ \bibinfo {pages}
  {1482} (\bibinfo {year} {2008})}\BibitemShut {NoStop}%
\bibitem [{\citenamefont {Huismans}\ \emph {et~al.}(2011)\citenamefont
  {Huismans}, \citenamefont {Rouz{\'{e}}e}, \citenamefont {Gijsbertsen},
  \citenamefont {Jungmann}, \citenamefont {Smolkowska}, \citenamefont {Logman},
  \citenamefont {L{\'{e}}pine}, \citenamefont {Cauchy}, \citenamefont {Zamith},
  \citenamefont {Marchenko}, \citenamefont {Bakker}, \citenamefont {Berden},
  \citenamefont {Redlich}, \citenamefont {van~der Meer}, \citenamefont
  {Muller}, \citenamefont {Vermin}, \citenamefont {Schafer}, \citenamefont
  {Spanner}, \citenamefont {Ivanov}, \citenamefont {Smirnova}, \citenamefont
  {Bauer}, \citenamefont {Popruzhenko},\ and\ \citenamefont
  {Vrakking}}]{Huismans_2011}%
  \BibitemOpen
  \bibfield  {author} {\bibinfo {author} {\bibfnamefont {Y.}~\bibnamefont
  {Huismans}}, \bibinfo {author} {\bibfnamefont {A.}~\bibnamefont
  {Rouz{\'{e}}e}}, \bibinfo {author} {\bibfnamefont {A.}~\bibnamefont
  {Gijsbertsen}}, \bibinfo {author} {\bibfnamefont {J.~H.}\ \bibnamefont
  {Jungmann}}, \bibinfo {author} {\bibfnamefont {A.~S.}\ \bibnamefont
  {Smolkowska}}, \bibinfo {author} {\bibfnamefont {P.~S. W.~M.}\ \bibnamefont
  {Logman}}, \bibinfo {author} {\bibfnamefont {F.}~\bibnamefont
  {L{\'{e}}pine}}, \bibinfo {author} {\bibfnamefont {C.}~\bibnamefont
  {Cauchy}}, \bibinfo {author} {\bibfnamefont {S.}~\bibnamefont {Zamith}},
  \bibinfo {author} {\bibfnamefont {T.}~\bibnamefont {Marchenko}}, \bibinfo
  {author} {\bibfnamefont {J.~M.}\ \bibnamefont {Bakker}}, \bibinfo {author}
  {\bibfnamefont {G.}~\bibnamefont {Berden}}, \bibinfo {author} {\bibfnamefont
  {B.}~\bibnamefont {Redlich}}, \bibinfo {author} {\bibfnamefont {A.~F.~G.}\
  \bibnamefont {van~der Meer}}, \bibinfo {author} {\bibfnamefont {H.~G.}\
  \bibnamefont {Muller}}, \bibinfo {author} {\bibfnamefont {W.}~\bibnamefont
  {Vermin}}, \bibinfo {author} {\bibfnamefont {K.~J.}\ \bibnamefont {Schafer}},
  \bibinfo {author} {\bibfnamefont {M.}~\bibnamefont {Spanner}}, \bibinfo
  {author} {\bibfnamefont {M.~Y.}\ \bibnamefont {Ivanov}}, \bibinfo {author}
  {\bibfnamefont {O.}~\bibnamefont {Smirnova}}, \bibinfo {author}
  {\bibfnamefont {D.}~\bibnamefont {Bauer}}, \bibinfo {author} {\bibfnamefont
  {S.~V.}\ \bibnamefont {Popruzhenko}},\ and\ \bibinfo {author} {\bibfnamefont
  {M.~J.~J.}\ \bibnamefont {Vrakking}},\ }\bibfield  {title} {\bibinfo {title}
  {{Time-resolved holography with photoelectrons}},\ }\href
  {https://doi.org/10.1126/science.1198450} {\bibfield  {journal} {\bibinfo
  {journal} {Science (New York, N.Y.)}\ }\textbf {\bibinfo {volume} {331}},\
  \bibinfo {pages} {61} (\bibinfo {year} {2011})}\BibitemShut {NoStop}%
\bibitem [{\citenamefont {Agostini}(2024)}]{Agostini_2024}%
  \BibitemOpen
  \bibfield  {author} {\bibinfo {author} {\bibfnamefont {P.}~\bibnamefont
  {Agostini}},\ }\bibfield  {title} {\bibinfo {title} {Nobel lecture: Genesis
  and applications of attosecond pulse trains},\ }\href
  {https://doi.org/10.1103/RevModPhys.96.030501} {\bibfield  {journal}
  {\bibinfo  {journal} {Rev. Mod. Phys.}\ }\textbf {\bibinfo {volume} {96}},\
  \bibinfo {pages} {030501} (\bibinfo {year} {2024})}\BibitemShut {NoStop}%
\bibitem [{\citenamefont {L'Huillier}(2024)}]{L'Huillier_2024}%
  \BibitemOpen
  \bibfield  {author} {\bibinfo {author} {\bibfnamefont {A.}~\bibnamefont
  {L'Huillier}},\ }\bibfield  {title} {\bibinfo {title} {Nobel lecture: The
  route to attosecond pulses},\ }\href
  {https://doi.org/10.1103/RevModPhys.96.030503} {\bibfield  {journal}
  {\bibinfo  {journal} {Rev. Mod. Phys.}\ }\textbf {\bibinfo {volume} {96}},\
  \bibinfo {pages} {030503} (\bibinfo {year} {2024})}\BibitemShut {NoStop}%
\bibitem [{\citenamefont {Krausz}(2024)}]{Krausz_2024}%
  \BibitemOpen
  \bibfield  {author} {\bibinfo {author} {\bibfnamefont {F.}~\bibnamefont
  {Krausz}},\ }\bibfield  {title} {\bibinfo {title} {Nobel lecture: Sub-atomic
  motions},\ }\href {https://doi.org/10.1103/RevModPhys.96.030502} {\bibfield
  {journal} {\bibinfo  {journal} {Rev. Mod. Phys.}\ }\textbf {\bibinfo {volume}
  {96}},\ \bibinfo {pages} {030502} (\bibinfo {year} {2024})}\BibitemShut
  {NoStop}%
\bibitem [{\citenamefont {Brabec}\ \emph {et~al.}(1996)\citenamefont {Brabec},
  \citenamefont {Ivanov},\ and\ \citenamefont {Corkum}}]{Brabec_1996}%
  \BibitemOpen
  \bibfield  {author} {\bibinfo {author} {\bibfnamefont {T.}~\bibnamefont
  {Brabec}}, \bibinfo {author} {\bibfnamefont {M.~Y.}\ \bibnamefont {Ivanov}},\
  and\ \bibinfo {author} {\bibfnamefont {P.~B.}\ \bibnamefont {Corkum}},\
  }\bibfield  {title} {\bibinfo {title} {Coulomb focusing in intense field
  atomic processes},\ }\href@noop {} {\bibfield  {journal} {\bibinfo  {journal}
  {Phys. Rev. A}\ }\textbf {\bibinfo {volume} {54}},\ \bibinfo {pages} {R2551}
  (\bibinfo {year} {1996})}\BibitemShut {NoStop}%
\bibitem [{\citenamefont {Comtois}\ \emph {et~al.}(2005)\citenamefont
  {Comtois}, \citenamefont {Zeidler}, \citenamefont {P\'{e}pin}, \citenamefont
  {Kieffer}, \citenamefont {Villeneuve},\ and\ \citenamefont
  {Corkum}}]{Comtois_2005}%
  \BibitemOpen
  \bibfield  {author} {\bibinfo {author} {\bibfnamefont {D.}~\bibnamefont
  {Comtois}}, \bibinfo {author} {\bibfnamefont {D.}~\bibnamefont {Zeidler}},
  \bibinfo {author} {\bibfnamefont {H.}~\bibnamefont {P\'{e}pin}}, \bibinfo
  {author} {\bibfnamefont {J.~C.}\ \bibnamefont {Kieffer}}, \bibinfo {author}
  {\bibfnamefont {D.~M.}\ \bibnamefont {Villeneuve}},\ and\ \bibinfo {author}
  {\bibfnamefont {P.~B.}\ \bibnamefont {Corkum}},\ }\bibfield  {title}
  {\bibinfo {title} {Observation of coulomb focusing in tunnelling ionization
  of noble gases},\ }\href@noop {} {\bibfield  {journal} {\bibinfo  {journal}
  {J. Phys. B}\ }\textbf {\bibinfo {volume} {38}},\ \bibinfo {pages} {1923}
  (\bibinfo {year} {2005})}\BibitemShut {NoStop}%
\bibitem [{\citenamefont {Yudin}\ and\ \citenamefont
  {Ivanov}(2001)}]{Yudin_2001a}%
  \BibitemOpen
  \bibfield  {author} {\bibinfo {author} {\bibfnamefont {G.~L.}\ \bibnamefont
  {Yudin}}\ and\ \bibinfo {author} {\bibfnamefont {M.~Y.}\ \bibnamefont
  {Ivanov}},\ }\bibfield  {title} {\bibinfo {title} {Physics of correlated
  double ionization of atoms in intense laser fields: Quasistatic tunneling
  limit},\ }\href@noop {} {\bibfield  {journal} {\bibinfo  {journal} {Phys.
  Rev. A}\ }\textbf {\bibinfo {volume} {63}},\ \bibinfo {pages} {033404}
  (\bibinfo {year} {2001})}\BibitemShut {NoStop}%
\bibitem [{\citenamefont {Rudenko}\ \emph {et~al.}(2005)\citenamefont
  {Rudenko}, \citenamefont {Zrost}, \citenamefont {Ergler}, \citenamefont
  {Voitkiv}, \citenamefont {Najjari}, \citenamefont {de~Jesus}, \citenamefont
  {Feuerstein}, \citenamefont {Schr\"{o}ter}, \citenamefont {Moshammer},\ and\
  \citenamefont {Ullrich}}]{Rudenko_2005}%
  \BibitemOpen
  \bibfield  {author} {\bibinfo {author} {\bibfnamefont {A.}~\bibnamefont
  {Rudenko}}, \bibinfo {author} {\bibfnamefont {K.}~\bibnamefont {Zrost}},
  \bibinfo {author} {\bibfnamefont {T.}~\bibnamefont {Ergler}}, \bibinfo
  {author} {\bibfnamefont {A.~B.}\ \bibnamefont {Voitkiv}}, \bibinfo {author}
  {\bibfnamefont {B.}~\bibnamefont {Najjari}}, \bibinfo {author} {\bibfnamefont
  {V.~L.~B.}\ \bibnamefont {de~Jesus}}, \bibinfo {author} {\bibfnamefont
  {B.}~\bibnamefont {Feuerstein}}, \bibinfo {author} {\bibfnamefont {C.~D.}\
  \bibnamefont {Schr\"{o}ter}}, \bibinfo {author} {\bibfnamefont
  {R.}~\bibnamefont {Moshammer}},\ and\ \bibinfo {author} {\bibfnamefont
  {J.}~\bibnamefont {Ullrich}},\ }\bibfield  {title} {\bibinfo {title} {Coulomb
  singularity in the transverse momentum distribution for strong-field single
  ionization},\ }\href@noop {} {\bibfield  {journal} {\bibinfo  {journal} {J.
  Phys. B}\ }\textbf {\bibinfo {volume} {38}},\ \bibinfo {pages} {L191}
  (\bibinfo {year} {2005})}\BibitemShut {NoStop}%
\bibitem [{\citenamefont {Ludwig}\ \emph {et~al.}(2014)\citenamefont {Ludwig},
  \citenamefont {Maurer}, \citenamefont {Mayer}, \citenamefont {Phillips},
  \citenamefont {Gallmann},\ and\ \citenamefont {Keller}}]{Ludwig_2014}%
  \BibitemOpen
  \bibfield  {author} {\bibinfo {author} {\bibfnamefont {A.}~\bibnamefont
  {Ludwig}}, \bibinfo {author} {\bibfnamefont {J.}~\bibnamefont {Maurer}},
  \bibinfo {author} {\bibfnamefont {B.~W.}\ \bibnamefont {Mayer}}, \bibinfo
  {author} {\bibfnamefont {C.~R.}\ \bibnamefont {Phillips}}, \bibinfo {author}
  {\bibfnamefont {L.}~\bibnamefont {Gallmann}},\ and\ \bibinfo {author}
  {\bibfnamefont {U.}~\bibnamefont {Keller}},\ }\bibfield  {title} {\bibinfo
  {title} {Breakdown of the dipole approximation in strong-field ionization},\
  }\href@noop {} {\bibfield  {journal} {\bibinfo  {journal} {Phys. Rev. Lett.}\
  }\textbf {\bibinfo {volume} {113}},\ \bibinfo {pages} {243001} (\bibinfo
  {year} {2014})}\BibitemShut {NoStop}%
\bibitem [{\citenamefont {Dan\v{e}k}\ \emph {et~al.}(2018)\citenamefont
  {Dan\v{e}k}, \citenamefont {Hatsagortsyan},\ and\ \citenamefont
  {Keitel}}]{Danek_2018}%
  \BibitemOpen
  \bibfield  {author} {\bibinfo {author} {\bibfnamefont {J.}~\bibnamefont
  {Dan\v{e}k}}, \bibinfo {author} {\bibfnamefont {K.~Z.}\ \bibnamefont
  {Hatsagortsyan}},\ and\ \bibinfo {author} {\bibfnamefont {C.~H.}\
  \bibnamefont {Keitel}},\ }\bibfield  {title} {\bibinfo {title} {{Analytical
  approach to Coulomb focusing in strong-field ionization. I. Nondipole
  effects}},\ }\href@noop {} {\bibfield  {journal} {\bibinfo  {journal} {Phys.
  Rev. A}\ }\textbf {\bibinfo {volume} {97}},\ \bibinfo {pages} {063409}
  (\bibinfo {year} {2018})}\BibitemShut {NoStop}%
\bibitem [{\citenamefont {Willenberg}\ \emph {et~al.}(2019)\citenamefont
  {Willenberg}, \citenamefont {Maurer}, \citenamefont {Keller}, \citenamefont
  {Dan\ifmmode~\check{e}\else \v{e}\fi{}k}, \citenamefont {Klaiber},
  \citenamefont {Teeny}, \citenamefont {Hatsagortsyan},\ and\ \citenamefont
  {Keitel}}]{Willenberg_2019}%
  \BibitemOpen
  \bibfield  {author} {\bibinfo {author} {\bibfnamefont {B.}~\bibnamefont
  {Willenberg}}, \bibinfo {author} {\bibfnamefont {J.}~\bibnamefont {Maurer}},
  \bibinfo {author} {\bibfnamefont {U.}~\bibnamefont {Keller}}, \bibinfo
  {author} {\bibfnamefont {J.}~\bibnamefont {Dan\ifmmode~\check{e}\else
  \v{e}\fi{}k}}, \bibinfo {author} {\bibfnamefont {M.}~\bibnamefont {Klaiber}},
  \bibinfo {author} {\bibfnamefont {N.}~\bibnamefont {Teeny}}, \bibinfo
  {author} {\bibfnamefont {K.~Z.}\ \bibnamefont {Hatsagortsyan}},\ and\
  \bibinfo {author} {\bibfnamefont {C.~H.}\ \bibnamefont {Keitel}},\ }\bibfield
   {title} {\bibinfo {title} {Holographic interferences in strong-field
  ionization beyond the dipole approximation: The influence of the peak and
  focal-volume-averaged laser intensities},\ }\href
  {https://doi.org/10.1103/PhysRevA.100.033417} {\bibfield  {journal} {\bibinfo
   {journal} {Phys. Rev. A}\ }\textbf {\bibinfo {volume} {100}},\ \bibinfo
  {pages} {033417} (\bibinfo {year} {2019})}\BibitemShut {NoStop}%
\bibitem [{\citenamefont {Blaga}\ \emph {et~al.}(2009)\citenamefont {Blaga},
  \citenamefont {Catoire}, \citenamefont {Colosimo}, \citenamefont {Paulus},
  \citenamefont {Muller}, \citenamefont {P.},\ and\ \citenamefont
  {DiMauro}}]{Blaga_2009}%
  \BibitemOpen
  \bibfield  {author} {\bibinfo {author} {\bibfnamefont {C.~I.}\ \bibnamefont
  {Blaga}}, \bibinfo {author} {\bibfnamefont {F.}~\bibnamefont {Catoire}},
  \bibinfo {author} {\bibfnamefont {P.}~\bibnamefont {Colosimo}}, \bibinfo
  {author} {\bibfnamefont {G.~G.}\ \bibnamefont {Paulus}}, \bibinfo {author}
  {\bibfnamefont {H.~G.}\ \bibnamefont {Muller}}, \bibinfo {author}
  {\bibfnamefont {A.}~\bibnamefont {P.}},\ and\ \bibinfo {author}
  {\bibfnamefont {L.~F.}\ \bibnamefont {DiMauro}},\ }\bibfield  {title}
  {\bibinfo {title} {Strong-field photoionization revisited},\ }\href@noop {}
  {\bibfield  {journal} {\bibinfo  {journal} {Nat. Phys.}\ }\textbf {\bibinfo
  {volume} {5}},\ \bibinfo {pages} {1745} (\bibinfo {year} {2009})}\BibitemShut
  {NoStop}%
\bibitem [{\citenamefont {Faisal}(2009)}]{Faisal_2009}%
  \BibitemOpen
  \bibfield  {author} {\bibinfo {author} {\bibfnamefont {F.~H.~M.}\
  \bibnamefont {Faisal}},\ }\bibfield  {title} {\bibinfo {title} {{Strong-field
  physics: Ionization surprise}},\ }\href@noop {} {\bibfield  {journal}
  {\bibinfo  {journal} {Nat. Phys.}\ }\textbf {\bibinfo {volume} {5}},\
  \bibinfo {pages} {319} (\bibinfo {year} {2009})}\BibitemShut {NoStop}%
\bibitem [{\citenamefont {Wu}\ \emph {et~al.}(2012)\citenamefont {Wu},
  \citenamefont {Yang}, \citenamefont {Liu}, \citenamefont {Gong},
  \citenamefont {Wu}, \citenamefont {Liu}, \citenamefont {Hao}, \citenamefont
  {Li}, \citenamefont {He},\ and\ \citenamefont {Chen}}]{Wu_2012b}%
  \BibitemOpen
  \bibfield  {author} {\bibinfo {author} {\bibfnamefont {C.~Y.}\ \bibnamefont
  {Wu}}, \bibinfo {author} {\bibfnamefont {Y.~D.}\ \bibnamefont {Yang}},
  \bibinfo {author} {\bibfnamefont {Y.~Q.}\ \bibnamefont {Liu}}, \bibinfo
  {author} {\bibfnamefont {Q.~H.}\ \bibnamefont {Gong}}, \bibinfo {author}
  {\bibfnamefont {M.}~\bibnamefont {Wu}}, \bibinfo {author} {\bibfnamefont
  {X.}~\bibnamefont {Liu}}, \bibinfo {author} {\bibfnamefont {X.~L.}\
  \bibnamefont {Hao}}, \bibinfo {author} {\bibfnamefont {W.~D.}\ \bibnamefont
  {Li}}, \bibinfo {author} {\bibfnamefont {X.~T.}\ \bibnamefont {He}},\ and\
  \bibinfo {author} {\bibfnamefont {J.}~\bibnamefont {Chen}},\ }\bibfield
  {title} {\bibinfo {title} {Characteristic spectrum of very low-energy
  photoelectron from above-threshold ionization in the tunneling regime},\
  }\href@noop {} {\bibfield  {journal} {\bibinfo  {journal} {Phys. Rev. Lett.}\
  }\textbf {\bibinfo {volume} {109}},\ \bibinfo {pages} {043001} (\bibinfo
  {year} {2012})}\BibitemShut {NoStop}%
\bibitem [{\citenamefont {Liu}\ and\ \citenamefont
  {Hatsagortsyan}(2010)}]{Liu_2010}%
  \BibitemOpen
  \bibfield  {author} {\bibinfo {author} {\bibfnamefont {C.}~\bibnamefont
  {Liu}}\ and\ \bibinfo {author} {\bibfnamefont {K.~Z.}\ \bibnamefont
  {Hatsagortsyan}},\ }\bibfield  {title} {\bibinfo {title} {Origin of
  unexpected low energy structure in photoelectron spectra induced by
  midinfrared strong laser fields},\ }\href@noop {} {\bibfield  {journal}
  {\bibinfo  {journal} {Phys. Rev. Lett.}\ }\textbf {\bibinfo {volume} {105}},\
  \bibinfo {pages} {113003} (\bibinfo {year} {2010})}\BibitemShut {NoStop}%
\bibitem [{\citenamefont {Yan}\ \emph {et~al.}(2010)\citenamefont {Yan},
  \citenamefont {Popruzhenko}, \citenamefont {Vrakking},\ and\ \citenamefont
  {Bauer}}]{Yan_2010}%
  \BibitemOpen
  \bibfield  {author} {\bibinfo {author} {\bibfnamefont {T.-M.}\ \bibnamefont
  {Yan}}, \bibinfo {author} {\bibfnamefont {S.~V.}\ \bibnamefont
  {Popruzhenko}}, \bibinfo {author} {\bibfnamefont {M.~J.~J.}\ \bibnamefont
  {Vrakking}},\ and\ \bibinfo {author} {\bibfnamefont {D.}~\bibnamefont
  {Bauer}},\ }\bibfield  {title} {\bibinfo {title} {Low-energy structures in
  strong field ionization revealed by quantum orbits},\ }\href@noop {}
  {\bibfield  {journal} {\bibinfo  {journal} {Phys. Rev. Lett.}\ }\textbf
  {\bibinfo {volume} {105}},\ \bibinfo {pages} {253002} (\bibinfo {year}
  {2010})}\BibitemShut {NoStop}%
\bibitem [{\citenamefont {K\"astner}\ \emph {et~al.}(2012)\citenamefont
  {K\"astner}, \citenamefont {Saalmann},\ and\ \citenamefont
  {Rost}}]{Kastner_2012}%
  \BibitemOpen
  \bibfield  {author} {\bibinfo {author} {\bibfnamefont {A.}~\bibnamefont
  {K\"astner}}, \bibinfo {author} {\bibfnamefont {U.}~\bibnamefont
  {Saalmann}},\ and\ \bibinfo {author} {\bibfnamefont {J.~M.}\ \bibnamefont
  {Rost}},\ }\bibfield  {title} {\bibinfo {title} {Electron-energy bunching in
  laser-driven soft recollisions},\ }\href@noop {} {\bibfield  {journal}
  {\bibinfo  {journal} {Phys. Rev. Lett.}\ }\textbf {\bibinfo {volume} {108}},\
  \bibinfo {pages} {033201} (\bibinfo {year} {2012})}\BibitemShut {NoStop}%
\bibitem [{\citenamefont {Lemell}\ \emph {et~al.}(2012)\citenamefont {Lemell},
  \citenamefont {Dimitriou}, \citenamefont {Tong}, \citenamefont {Nagele},
  \citenamefont {Kartashov}, \citenamefont {Burgd\"orfer},\ and\ \citenamefont
  {Gr\"afe}}]{Lemell_2012}%
  \BibitemOpen
  \bibfield  {author} {\bibinfo {author} {\bibfnamefont {C.}~\bibnamefont
  {Lemell}}, \bibinfo {author} {\bibfnamefont {K.~I.}\ \bibnamefont
  {Dimitriou}}, \bibinfo {author} {\bibfnamefont {X.-M.}\ \bibnamefont {Tong}},
  \bibinfo {author} {\bibfnamefont {S.}~\bibnamefont {Nagele}}, \bibinfo
  {author} {\bibfnamefont {D.~V.}\ \bibnamefont {Kartashov}}, \bibinfo {author}
  {\bibfnamefont {J.}~\bibnamefont {Burgd\"orfer}},\ and\ \bibinfo {author}
  {\bibfnamefont {S.}~\bibnamefont {Gr\"afe}},\ }\bibfield  {title} {\bibinfo
  {title} {Low-energy peak structure in strong-field ionization by midinfrared
  laser pulses: Two-dimensional focusing by the atomic potential},\ }\href@noop
  {} {\bibfield  {journal} {\bibinfo  {journal} {Phys. Rev. A}\ }\textbf
  {\bibinfo {volume} {85}},\ \bibinfo {pages} {011403} (\bibinfo {year}
  {2012})}\BibitemShut {NoStop}%
\bibitem [{\citenamefont {Becker}\ \emph {et~al.}(2014)\citenamefont {Becker},
  \citenamefont {Goreslavski}, \citenamefont {Milo{\v{s}}evi{\'{c}}},\ and\
  \citenamefont {Paulus}}]{Becker_2014}%
  \BibitemOpen
  \bibfield  {author} {\bibinfo {author} {\bibfnamefont {W.}~\bibnamefont
  {Becker}}, \bibinfo {author} {\bibfnamefont {S.~P.}\ \bibnamefont
  {Goreslavski}}, \bibinfo {author} {\bibfnamefont {D.~B.}\ \bibnamefont
  {Milo{\v{s}}evi{\'{c}}}},\ and\ \bibinfo {author} {\bibfnamefont {G.~G.}\
  \bibnamefont {Paulus}},\ }\bibfield  {title} {\bibinfo {title} {{Low-energy
  electron rescattering in laser-induced ionization}},\ }\href@noop {}
  {\bibfield  {journal} {\bibinfo  {journal} {J. Phys. B}\ }\textbf {\bibinfo
  {volume} {47}},\ \bibinfo {pages} {204022} (\bibinfo {year}
  {2014})}\BibitemShut {NoStop}%
\bibitem [{\citenamefont {Wolter}\ \emph {et~al.}(2015)\citenamefont {Wolter},
  \citenamefont {Pullen}, \citenamefont {Baudisch}, \citenamefont {Sclafani},
  \citenamefont {Hemmer}, \citenamefont {Senftleben}, \citenamefont
  {Schr\"oter}, \citenamefont {Ullrich}, \citenamefont {Moshammer},\ and\
  \citenamefont {Biegert}}]{Wolter_2015x}%
  \BibitemOpen
  \bibfield  {author} {\bibinfo {author} {\bibfnamefont {B.}~\bibnamefont
  {Wolter}}, \bibinfo {author} {\bibfnamefont {M.~G.}\ \bibnamefont {Pullen}},
  \bibinfo {author} {\bibfnamefont {M.}~\bibnamefont {Baudisch}}, \bibinfo
  {author} {\bibfnamefont {M.}~\bibnamefont {Sclafani}}, \bibinfo {author}
  {\bibfnamefont {M.}~\bibnamefont {Hemmer}}, \bibinfo {author} {\bibfnamefont
  {A.}~\bibnamefont {Senftleben}}, \bibinfo {author} {\bibfnamefont {C.~D.}\
  \bibnamefont {Schr\"oter}}, \bibinfo {author} {\bibfnamefont
  {J.}~\bibnamefont {Ullrich}}, \bibinfo {author} {\bibfnamefont
  {R.}~\bibnamefont {Moshammer}},\ and\ \bibinfo {author} {\bibfnamefont
  {J.}~\bibnamefont {Biegert}},\ }\bibfield  {title} {\bibinfo {title}
  {Strong-field physics with mid-ir fields},\ }\href@noop {} {\bibfield
  {journal} {\bibinfo  {journal} {Phys. Rev. X}\ }\textbf {\bibinfo {volume}
  {5}},\ \bibinfo {pages} {021034} (\bibinfo {year} {2015})}\BibitemShut
  {NoStop}%
\bibitem [{\citenamefont {Paulus}\ \emph {et~al.}(1998)\citenamefont {Paulus},
  \citenamefont {Zacher}, \citenamefont {Walther}, \citenamefont {Lohr},
  \citenamefont {Becker},\ and\ \citenamefont {Kleber}}]{Paulus_1998}%
  \BibitemOpen
  \bibfield  {author} {\bibinfo {author} {\bibfnamefont {G.~G.}\ \bibnamefont
  {Paulus}}, \bibinfo {author} {\bibfnamefont {F.}~\bibnamefont {Zacher}},
  \bibinfo {author} {\bibfnamefont {H.}~\bibnamefont {Walther}}, \bibinfo
  {author} {\bibfnamefont {A.}~\bibnamefont {Lohr}}, \bibinfo {author}
  {\bibfnamefont {W.}~\bibnamefont {Becker}},\ and\ \bibinfo {author}
  {\bibfnamefont {M.}~\bibnamefont {Kleber}},\ }\bibfield  {title} {\bibinfo
  {title} {Above-threshold ionization by an elliptically polarized field:
  Quantum tunneling interferences and classical dodging},\ }\href
  {https://doi.org/10.1103/PhysRevLett.80.484} {\bibfield  {journal} {\bibinfo
  {journal} {Phys. Rev. Lett.}\ }\textbf {\bibinfo {volume} {80}},\ \bibinfo
  {pages} {484} (\bibinfo {year} {1998})}\BibitemShut {NoStop}%
\bibitem [{\citenamefont {Kopold}\ \emph {et~al.}(2000)\citenamefont {Kopold},
  \citenamefont {Milo\ifmmode \check{s}\else
  \v{s}\fi{}evi\ifmmode~\acute{c}\else \'{c}\fi{}},\ and\ \citenamefont
  {Becker}}]{Kopold_2000}%
  \BibitemOpen
  \bibfield  {author} {\bibinfo {author} {\bibfnamefont {R.}~\bibnamefont
  {Kopold}}, \bibinfo {author} {\bibfnamefont {D.~B.}\ \bibnamefont
  {Milo\ifmmode \check{s}\else \v{s}\fi{}evi\ifmmode~\acute{c}\else
  \'{c}\fi{}}},\ and\ \bibinfo {author} {\bibfnamefont {W.}~\bibnamefont
  {Becker}},\ }\bibfield  {title} {\bibinfo {title} {Rescattering processes for
  elliptical polarization: A quantum trajectory analysis},\ }\href
  {https://doi.org/10.1103/PhysRevLett.84.3831} {\bibfield  {journal} {\bibinfo
   {journal} {Phys. Rev. Lett.}\ }\textbf {\bibinfo {volume} {84}},\ \bibinfo
  {pages} {3831} (\bibinfo {year} {2000})}\BibitemShut {NoStop}%
\bibitem [{\citenamefont {Paulus}\ \emph {et~al.}(2000)\citenamefont {Paulus},
  \citenamefont {Grasbon}, \citenamefont {Dreischuh}, \citenamefont {Walther},
  \citenamefont {Kopold},\ and\ \citenamefont {Becker}}]{Paulus_2000}%
  \BibitemOpen
  \bibfield  {author} {\bibinfo {author} {\bibfnamefont {G.~G.}\ \bibnamefont
  {Paulus}}, \bibinfo {author} {\bibfnamefont {F.}~\bibnamefont {Grasbon}},
  \bibinfo {author} {\bibfnamefont {A.}~\bibnamefont {Dreischuh}}, \bibinfo
  {author} {\bibfnamefont {H.}~\bibnamefont {Walther}}, \bibinfo {author}
  {\bibfnamefont {R.}~\bibnamefont {Kopold}},\ and\ \bibinfo {author}
  {\bibfnamefont {W.}~\bibnamefont {Becker}},\ }\bibfield  {title} {\bibinfo
  {title} {Above-threshold ionization by an elliptically polarized field:
  Interplay between electronic quantum trajectories},\ }\href
  {https://doi.org/10.1103/PhysRevLett.84.3791} {\bibfield  {journal} {\bibinfo
   {journal} {Phys. Rev. Lett.}\ }\textbf {\bibinfo {volume} {84}},\ \bibinfo
  {pages} {3791} (\bibinfo {year} {2000})}\BibitemShut {NoStop}%
\bibitem [{\citenamefont {M\"oller}\ \emph {et~al.}(2012)\citenamefont
  {M\"oller}, \citenamefont {Cheng}, \citenamefont {Khan}, \citenamefont
  {Zhao}, \citenamefont {Zhao}, \citenamefont {Chini}, \citenamefont {Paulus},\
  and\ \citenamefont {Chang}}]{Moeller_2012}%
  \BibitemOpen
  \bibfield  {author} {\bibinfo {author} {\bibfnamefont {M.}~\bibnamefont
  {M\"oller}}, \bibinfo {author} {\bibfnamefont {Y.}~\bibnamefont {Cheng}},
  \bibinfo {author} {\bibfnamefont {S.~D.}\ \bibnamefont {Khan}}, \bibinfo
  {author} {\bibfnamefont {B.}~\bibnamefont {Zhao}}, \bibinfo {author}
  {\bibfnamefont {K.}~\bibnamefont {Zhao}}, \bibinfo {author} {\bibfnamefont
  {M.}~\bibnamefont {Chini}}, \bibinfo {author} {\bibfnamefont {G.~G.}\
  \bibnamefont {Paulus}},\ and\ \bibinfo {author} {\bibfnamefont
  {Z.}~\bibnamefont {Chang}},\ }\bibfield  {title} {\bibinfo {title}
  {Dependence of high-order-harmonic-generation yield on driving-laser
  ellipticity},\ }\href {https://doi.org/10.1103/PhysRevA.86.011401} {\bibfield
   {journal} {\bibinfo  {journal} {Phys. Rev. A}\ }\textbf {\bibinfo {volume}
  {86}},\ \bibinfo {pages} {011401} (\bibinfo {year} {2012})}\BibitemShut
  {NoStop}%
\bibitem [{\citenamefont {Lai}\ \emph {et~al.}(2013)\citenamefont {Lai},
  \citenamefont {Wang}, \citenamefont {Chen}, \citenamefont {Hu}, \citenamefont
  {Quan}, \citenamefont {Liu}, \citenamefont {Chen}, \citenamefont {Cheng},
  \citenamefont {Xu},\ and\ \citenamefont {Becker}}]{Lai_2013}%
  \BibitemOpen
  \bibfield  {author} {\bibinfo {author} {\bibfnamefont {X.}~\bibnamefont
  {Lai}}, \bibinfo {author} {\bibfnamefont {C.}~\bibnamefont {Wang}}, \bibinfo
  {author} {\bibfnamefont {Y.}~\bibnamefont {Chen}}, \bibinfo {author}
  {\bibfnamefont {Z.}~\bibnamefont {Hu}}, \bibinfo {author} {\bibfnamefont
  {W.}~\bibnamefont {Quan}}, \bibinfo {author} {\bibfnamefont {X.}~\bibnamefont
  {Liu}}, \bibinfo {author} {\bibfnamefont {J.}~\bibnamefont {Chen}}, \bibinfo
  {author} {\bibfnamefont {Y.}~\bibnamefont {Cheng}}, \bibinfo {author}
  {\bibfnamefont {Z.}~\bibnamefont {Xu}},\ and\ \bibinfo {author}
  {\bibfnamefont {W.}~\bibnamefont {Becker}},\ }\bibfield  {title} {\bibinfo
  {title} {Elliptical polarization favors long quantum orbits in high-order
  above-threshold ionization of noble gases},\ }\href
  {https://doi.org/10.1103/PhysRevLett.110.043002} {\bibfield  {journal}
  {\bibinfo  {journal} {Phys. Rev. Lett.}\ }\textbf {\bibinfo {volume} {110}},\
  \bibinfo {pages} {043002} (\bibinfo {year} {2013})}\BibitemShut {NoStop}%
\bibitem [{\citenamefont {Shvetsov-Shilovski}\ \emph
  {et~al.}(2008)\citenamefont {Shvetsov-Shilovski}, \citenamefont
  {Goreslavski}, \citenamefont {Popruzhenko},\ and\ \citenamefont
  {Becker}}]{Shvetsov-Shilovski_2008}%
  \BibitemOpen
  \bibfield  {author} {\bibinfo {author} {\bibfnamefont {N.~I.}\ \bibnamefont
  {Shvetsov-Shilovski}}, \bibinfo {author} {\bibfnamefont {S.~P.}\ \bibnamefont
  {Goreslavski}}, \bibinfo {author} {\bibfnamefont {S.~V.}\ \bibnamefont
  {Popruzhenko}},\ and\ \bibinfo {author} {\bibfnamefont {W.}~\bibnamefont
  {Becker}},\ }\bibfield  {title} {\bibinfo {title} {Ellipticity effects and
  the contributions of long orbits in nonsequential double ionization of
  atoms},\ }\href@noop {} {\bibfield  {journal} {\bibinfo  {journal} {Phys.
  Rev. A}\ }\textbf {\bibinfo {volume} {77}},\ \bibinfo {pages} {063405}
  (\bibinfo {year} {2008})}\BibitemShut {NoStop}%
\bibitem [{\citenamefont {Wang}\ and\ \citenamefont
  {Eberly}(2009)}]{Wang_2009}%
  \BibitemOpen
  \bibfield  {author} {\bibinfo {author} {\bibfnamefont {X.}~\bibnamefont
  {Wang}}\ and\ \bibinfo {author} {\bibfnamefont {J.~H.}\ \bibnamefont
  {Eberly}},\ }\bibfield  {title} {\bibinfo {title} {Effects of elliptical
  polarization on strong-field short-pulse double ionization},\ }\href@noop {}
  {\bibfield  {journal} {\bibinfo  {journal} {Phys. Rev. Lett.}\ }\textbf
  {\bibinfo {volume} {103}},\ \bibinfo {pages} {103007} (\bibinfo {year}
  {2009})}\BibitemShut {NoStop}%
\bibitem [{\citenamefont {Wang}\ and\ \citenamefont
  {Eberly}(2010)}]{Wang_2010}%
  \BibitemOpen
  \bibfield  {author} {\bibinfo {author} {\bibfnamefont {X.}~\bibnamefont
  {Wang}}\ and\ \bibinfo {author} {\bibfnamefont {J.~H.}\ \bibnamefont
  {Eberly}},\ }\bibfield  {title} {\bibinfo {title} {Elliptical polarization
  and probability of double ionization},\ }\href@noop {} {\bibfield  {journal}
  {\bibinfo  {journal} {Phys. Rev. Lett.}\ }\textbf {\bibinfo {volume} {105}},\
  \bibinfo {pages} {083001} (\bibinfo {year} {2010})}\BibitemShut {NoStop}%
\bibitem [{\citenamefont {Keldysh}(1964)}]{Keldysh_1965}%
  \BibitemOpen
  \bibfield  {author} {\bibinfo {author} {\bibfnamefont {L.~V.}\ \bibnamefont
  {Keldysh}},\ }\bibfield  {title} {\bibinfo {title} {Ionization in the field
  of a strong electromagnetic wave},\ }\href@noop {} {\bibfield  {journal}
  {\bibinfo  {journal} {Zh. Eksp. Teor. Fiz.}\ }\textbf {\bibinfo {volume}
  {47}},\ \bibinfo {pages} {1945} (\bibinfo {year} {1964})}\BibitemShut
  {NoStop}%
\bibitem [{\citenamefont {Maurer}\ \emph {et~al.}(2018)\citenamefont {Maurer},
  \citenamefont {Willenberg}, \citenamefont {Dan\v{e}k}, \citenamefont {Mayer},
  \citenamefont {Phillips}, \citenamefont {Gallmann}, \citenamefont {Klaiber},
  \citenamefont {Hatsagortsyan}, \citenamefont {Keitel},\ and\ \citenamefont
  {Keller}}]{Maurer_2018}%
  \BibitemOpen
  \bibfield  {author} {\bibinfo {author} {\bibfnamefont {J.}~\bibnamefont
  {Maurer}}, \bibinfo {author} {\bibfnamefont {B.}~\bibnamefont {Willenberg}},
  \bibinfo {author} {\bibfnamefont {J.}~\bibnamefont {Dan\v{e}k}}, \bibinfo
  {author} {\bibfnamefont {B.~W.}\ \bibnamefont {Mayer}}, \bibinfo {author}
  {\bibfnamefont {C.~R.}\ \bibnamefont {Phillips}}, \bibinfo {author}
  {\bibfnamefont {L.}~\bibnamefont {Gallmann}}, \bibinfo {author}
  {\bibfnamefont {M.}~\bibnamefont {Klaiber}}, \bibinfo {author} {\bibfnamefont
  {K.~Z.}\ \bibnamefont {Hatsagortsyan}}, \bibinfo {author} {\bibfnamefont
  {C.~H.}\ \bibnamefont {Keitel}},\ and\ \bibinfo {author} {\bibfnamefont
  {U.}~\bibnamefont {Keller}},\ }\bibfield  {title} {\bibinfo {title} {Probing
  the ionization wave packet and recollision dynamics with an elliptically
  polarized strong laser field in the nondipole regime},\ }\href
  {https://doi.org/10.1103/PhysRevA.97.013404} {\bibfield  {journal} {\bibinfo
  {journal} {Phys. Rev. A}\ }\textbf {\bibinfo {volume} {97}},\ \bibinfo
  {pages} {013404} (\bibinfo {year} {2018})}\BibitemShut {NoStop}%
\bibitem [{\citenamefont {Liu}\ and\ \citenamefont
  {Hatsagortsyan}(2012)}]{Liu_2012les}%
  \BibitemOpen
  \bibfield  {author} {\bibinfo {author} {\bibfnamefont {C.}~\bibnamefont
  {Liu}}\ and\ \bibinfo {author} {\bibfnamefont {K.~Z.}\ \bibnamefont
  {Hatsagortsyan}},\ }\bibfield  {title} {\bibinfo {title} {{Coulomb focusing
  in above-threshold ionization in elliptically polarized midinfrared strong
  laser fields}},\ }\href@noop {} {\bibfield  {journal} {\bibinfo  {journal}
  {Phys. Rev. A}\ }\textbf {\bibinfo {volume} {85}},\ \bibinfo {pages} {023413}
  (\bibinfo {year} {2012})}\BibitemShut {NoStop}%
\bibitem [{\citenamefont {Shafir}\ \emph {et~al.}(2013)\citenamefont {Shafir},
  \citenamefont {Soifer}, \citenamefont {Vozzi}, \citenamefont {Johnson},
  \citenamefont {Hartung}, \citenamefont {Dube}, \citenamefont {Villeneuve},
  \citenamefont {Corkum}, \citenamefont {Dudovich},\ and\ \citenamefont
  {Staudte}}]{Shafir_2013}%
  \BibitemOpen
  \bibfield  {author} {\bibinfo {author} {\bibfnamefont {D.}~\bibnamefont
  {Shafir}}, \bibinfo {author} {\bibfnamefont {H.}~\bibnamefont {Soifer}},
  \bibinfo {author} {\bibfnamefont {C.}~\bibnamefont {Vozzi}}, \bibinfo
  {author} {\bibfnamefont {A.~S.}\ \bibnamefont {Johnson}}, \bibinfo {author}
  {\bibfnamefont {A.}~\bibnamefont {Hartung}}, \bibinfo {author} {\bibfnamefont
  {Z.}~\bibnamefont {Dube}}, \bibinfo {author} {\bibfnamefont {D.~M.}\
  \bibnamefont {Villeneuve}}, \bibinfo {author} {\bibfnamefont {P.~B.}\
  \bibnamefont {Corkum}}, \bibinfo {author} {\bibfnamefont {N.}~\bibnamefont
  {Dudovich}},\ and\ \bibinfo {author} {\bibfnamefont {A.}~\bibnamefont
  {Staudte}},\ }\bibfield  {title} {\bibinfo {title} {Trajectory-resolved
  coulomb focusing in tunnel ionization of atoms with intense, elliptically
  polarized laser pulses},\ }\href
  {https://doi.org/10.1103/PhysRevLett.111.023005} {\bibfield  {journal}
  {\bibinfo  {journal} {Phys. Rev. Lett.}\ }\textbf {\bibinfo {volume} {111}},\
  \bibinfo {pages} {023005} (\bibinfo {year} {2013})}\BibitemShut {NoStop}%
\bibitem [{\citenamefont {Landsman}\ \emph {et~al.}(2013)\citenamefont
  {Landsman}, \citenamefont {Hofmann}, \citenamefont {Pfeiffer}, \citenamefont
  {Cirelli},\ and\ \citenamefont {Keller}}]{Landsman_2013}%
  \BibitemOpen
  \bibfield  {author} {\bibinfo {author} {\bibfnamefont {A.~S.}\ \bibnamefont
  {Landsman}}, \bibinfo {author} {\bibfnamefont {C.}~\bibnamefont {Hofmann}},
  \bibinfo {author} {\bibfnamefont {A.~N.}\ \bibnamefont {Pfeiffer}}, \bibinfo
  {author} {\bibfnamefont {C.}~\bibnamefont {Cirelli}},\ and\ \bibinfo {author}
  {\bibfnamefont {U.}~\bibnamefont {Keller}},\ }\bibfield  {title} {\bibinfo
  {title} {{Unified Approach to Probing Coulomb Effects in Tunnel Ionization
  for Any Ellipticity of Laser Light}},\ }\href
  {https://doi.org/10.1103/PhysRevLett.111.263001} {\bibfield  {journal}
  {\bibinfo  {journal} {Phys. Rev. Lett.}\ }\textbf {\bibinfo {volume} {111}},\
  \bibinfo {pages} {263001} (\bibinfo {year} {2013})}\BibitemShut {NoStop}%
\bibitem [{\citenamefont {Li}\ \emph {et~al.}(2013)\citenamefont {Li},
  \citenamefont {Liu}, \citenamefont {Liu}, \citenamefont {Ning}, \citenamefont
  {Fu}, \citenamefont {Liu}, \citenamefont {Deng}, \citenamefont {Wu},
  \citenamefont {Peng},\ and\ \citenamefont {Gong}}]{Li_2013}%
  \BibitemOpen
  \bibfield  {author} {\bibinfo {author} {\bibfnamefont {M.}~\bibnamefont
  {Li}}, \bibinfo {author} {\bibfnamefont {Y.}~\bibnamefont {Liu}}, \bibinfo
  {author} {\bibfnamefont {H.}~\bibnamefont {Liu}}, \bibinfo {author}
  {\bibfnamefont {Q.}~\bibnamefont {Ning}}, \bibinfo {author} {\bibfnamefont
  {L.}~\bibnamefont {Fu}}, \bibinfo {author} {\bibfnamefont {J.}~\bibnamefont
  {Liu}}, \bibinfo {author} {\bibfnamefont {Y.}~\bibnamefont {Deng}}, \bibinfo
  {author} {\bibfnamefont {C.}~\bibnamefont {Wu}}, \bibinfo {author}
  {\bibfnamefont {L.-Y.}\ \bibnamefont {Peng}},\ and\ \bibinfo {author}
  {\bibfnamefont {Q.}~\bibnamefont {Gong}},\ }\bibfield  {title} {\bibinfo
  {title} {{Subcycle Dynamics of Coulomb Asymmetry in Strong Elliptical Laser
  Fields}},\ }\href {https://doi.org/10.1103/PhysRevLett.111.023006} {\bibfield
   {journal} {\bibinfo  {journal} {Phys. Rev. Lett.}\ }\textbf {\bibinfo
  {volume} {111}},\ \bibinfo {pages} {023006} (\bibinfo {year}
  {2013})}\BibitemShut {NoStop}%
\bibitem [{\citenamefont {Eckle}\ \emph
  {et~al.}(2008{\natexlab{a}})\citenamefont {Eckle}, \citenamefont {Smolarski},
  \citenamefont {Schlup}, \citenamefont {Biegert}, \citenamefont {Staudte},
  \citenamefont {Sch\"offler}, \citenamefont {Muller}, \citenamefont
  {D\"orner},\ and\ \citenamefont {Keller}}]{Eckle_2008a}%
  \BibitemOpen
  \bibfield  {author} {\bibinfo {author} {\bibfnamefont {P.}~\bibnamefont
  {Eckle}}, \bibinfo {author} {\bibfnamefont {M.}~\bibnamefont {Smolarski}},
  \bibinfo {author} {\bibfnamefont {F.}~\bibnamefont {Schlup}}, \bibinfo
  {author} {\bibfnamefont {J.}~\bibnamefont {Biegert}}, \bibinfo {author}
  {\bibfnamefont {A.}~\bibnamefont {Staudte}}, \bibinfo {author} {\bibfnamefont
  {M.}~\bibnamefont {Sch\"offler}}, \bibinfo {author} {\bibfnamefont {H.~G.}\
  \bibnamefont {Muller}}, \bibinfo {author} {\bibfnamefont {R.}~\bibnamefont
  {D\"orner}},\ and\ \bibinfo {author} {\bibfnamefont {U.}~\bibnamefont
  {Keller}},\ }\bibfield  {title} {\bibinfo {title} {Attosecond angular
  streaking},\ }\href@noop {} {\bibfield  {journal} {\bibinfo  {journal}
  {Nature Phys.}\ }\textbf {\bibinfo {volume} {4}},\ \bibinfo {pages} {565}
  (\bibinfo {year} {2008}{\natexlab{a}})}\BibitemShut {NoStop}%
\bibitem [{\citenamefont {Eckle}\ \emph
  {et~al.}(2008{\natexlab{b}})\citenamefont {Eckle}, \citenamefont {Pfeiffer},
  \citenamefont {Cirelli}, \citenamefont {Staudte}, \citenamefont {D\"orner},
  \citenamefont {Muller}, \citenamefont {B\"uttiker},\ and\ \citenamefont
  {Keller}}]{Eckle_2008b}%
  \BibitemOpen
  \bibfield  {author} {\bibinfo {author} {\bibfnamefont {P.}~\bibnamefont
  {Eckle}}, \bibinfo {author} {\bibfnamefont {A.~N.}\ \bibnamefont {Pfeiffer}},
  \bibinfo {author} {\bibfnamefont {C.}~\bibnamefont {Cirelli}}, \bibinfo
  {author} {\bibfnamefont {A.}~\bibnamefont {Staudte}}, \bibinfo {author}
  {\bibfnamefont {R.}~\bibnamefont {D\"orner}}, \bibinfo {author}
  {\bibfnamefont {H.~G.}\ \bibnamefont {Muller}}, \bibinfo {author}
  {\bibfnamefont {M.}~\bibnamefont {B\"uttiker}},\ and\ \bibinfo {author}
  {\bibfnamefont {U.}~\bibnamefont {Keller}},\ }\bibfield  {title} {\bibinfo
  {title} {Attosecond ionization and tunneling delay time measurements in
  helium},\ }\href@noop {} {\bibfield  {journal} {\bibinfo  {journal}
  {Science}\ }\textbf {\bibinfo {volume} {322}},\ \bibinfo {pages} {1525}
  (\bibinfo {year} {2008}{\natexlab{b}})}\BibitemShut {NoStop}%
\bibitem [{\citenamefont {Pfeiffer}\ \emph
  {et~al.}(2012{\natexlab{a}})\citenamefont {Pfeiffer}, \citenamefont
  {Cirelli}, \citenamefont {Smolarski}, \citenamefont {Dimitrovski},
  \citenamefont {Abu-samha}, \citenamefont {Madsen},\ and\ \citenamefont
  {Keller}}]{Pfeiffer_2012}%
  \BibitemOpen
  \bibfield  {author} {\bibinfo {author} {\bibfnamefont {A.~N.}\ \bibnamefont
  {Pfeiffer}}, \bibinfo {author} {\bibfnamefont {C.}~\bibnamefont {Cirelli}},
  \bibinfo {author} {\bibfnamefont {M.}~\bibnamefont {Smolarski}}, \bibinfo
  {author} {\bibfnamefont {D.}~\bibnamefont {Dimitrovski}}, \bibinfo {author}
  {\bibfnamefont {M.}~\bibnamefont {Abu-samha}}, \bibinfo {author}
  {\bibfnamefont {L.~B.}\ \bibnamefont {Madsen}},\ and\ \bibinfo {author}
  {\bibfnamefont {U.}~\bibnamefont {Keller}},\ }\bibfield  {title} {\bibinfo
  {title} {Attoclock reveals natural coordinates of the laser-induced
  tunnelling current flow in atoms},\ }\href@noop {} {\bibfield  {journal}
  {\bibinfo  {journal} {Nature Phys.}\ }\textbf {\bibinfo {volume} {8}},\
  \bibinfo {pages} {76} (\bibinfo {year} {2012}{\natexlab{a}})}\BibitemShut
  {NoStop}%
\bibitem [{\citenamefont {Landsman}\ \emph {et~al.}(2014)\citenamefont
  {Landsman}, \citenamefont {Weger}, \citenamefont {Maurer}, \citenamefont
  {Boge}, \citenamefont {Ludwig}, \citenamefont {Heuser}, \citenamefont
  {Cirelli}, \citenamefont {Gallmann},\ and\ \citenamefont
  {Keller}}]{Landsman_2014o}%
  \BibitemOpen
  \bibfield  {author} {\bibinfo {author} {\bibfnamefont {A.~S.}\ \bibnamefont
  {Landsman}}, \bibinfo {author} {\bibfnamefont {M.}~\bibnamefont {Weger}},
  \bibinfo {author} {\bibfnamefont {J.}~\bibnamefont {Maurer}}, \bibinfo
  {author} {\bibfnamefont {R.}~\bibnamefont {Boge}}, \bibinfo {author}
  {\bibfnamefont {A.}~\bibnamefont {Ludwig}}, \bibinfo {author} {\bibfnamefont
  {S.}~\bibnamefont {Heuser}}, \bibinfo {author} {\bibfnamefont
  {C.}~\bibnamefont {Cirelli}}, \bibinfo {author} {\bibfnamefont
  {L.}~\bibnamefont {Gallmann}},\ and\ \bibinfo {author} {\bibfnamefont
  {U.}~\bibnamefont {Keller}},\ }\bibfield  {title} {\bibinfo {title}
  {Ultrafast resolution of tunneling delay time},\ }\href@noop {} {\bibfield
  {journal} {\bibinfo  {journal} {Optica}\ }\textbf {\bibinfo {volume} {1}},\
  \bibinfo {pages} {343} (\bibinfo {year} {2014})}\BibitemShut {NoStop}%
\bibitem [{\citenamefont {Camus}\ \emph {et~al.}(2017)\citenamefont {Camus},
  \citenamefont {Yakaboylu}, \citenamefont {Fechner}, \citenamefont {Klaiber},
  \citenamefont {Laux}, \citenamefont {Mi}, \citenamefont {Hatsagortsyan},
  \citenamefont {Pfeifer}, \citenamefont {Keitel},\ and\ \citenamefont
  {Moshammer}}]{Camus_2017}%
  \BibitemOpen
  \bibfield  {author} {\bibinfo {author} {\bibfnamefont {N.}~\bibnamefont
  {Camus}}, \bibinfo {author} {\bibfnamefont {E.}~\bibnamefont {Yakaboylu}},
  \bibinfo {author} {\bibfnamefont {L.}~\bibnamefont {Fechner}}, \bibinfo
  {author} {\bibfnamefont {M.}~\bibnamefont {Klaiber}}, \bibinfo {author}
  {\bibfnamefont {M.}~\bibnamefont {Laux}}, \bibinfo {author} {\bibfnamefont
  {Y.}~\bibnamefont {Mi}}, \bibinfo {author} {\bibfnamefont {K.~Z.}\
  \bibnamefont {Hatsagortsyan}}, \bibinfo {author} {\bibfnamefont
  {T.}~\bibnamefont {Pfeifer}}, \bibinfo {author} {\bibfnamefont {C.~H.}\
  \bibnamefont {Keitel}},\ and\ \bibinfo {author} {\bibfnamefont
  {R.}~\bibnamefont {Moshammer}},\ }\bibfield  {title} {\bibinfo {title}
  {Experimental evidence for wigner's tunneling time},\ }\href@noop {}
  {\bibfield  {journal} {\bibinfo  {journal} {Phys. Rev. Lett.}\ }\textbf
  {\bibinfo {volume} {119}},\ \bibinfo {pages} {023201} (\bibinfo {year}
  {2017})}\BibitemShut {NoStop}%
\bibitem [{\citenamefont {Sainadh}\ \emph {et~al.}(2019)\citenamefont
  {Sainadh}, \citenamefont {Xu}, \citenamefont {Wang}, \citenamefont
  {Atia-Tul-Noor}, \citenamefont {Wallace}, \citenamefont {Douguet},
  \citenamefont {Bray}, \citenamefont {Ivanov}, \citenamefont {Bartschat},
  \citenamefont {Kheifets}, \citenamefont {Sang},\ and\ \citenamefont
  {Litvinyuk}}]{Sainadh_2019}%
  \BibitemOpen
  \bibfield  {author} {\bibinfo {author} {\bibfnamefont {U.~S.}\ \bibnamefont
  {Sainadh}}, \bibinfo {author} {\bibfnamefont {H.}~\bibnamefont {Xu}},
  \bibinfo {author} {\bibfnamefont {X.}~\bibnamefont {Wang}}, \bibinfo {author}
  {\bibfnamefont {A.}~\bibnamefont {Atia-Tul-Noor}}, \bibinfo {author}
  {\bibfnamefont {W.~C.}\ \bibnamefont {Wallace}}, \bibinfo {author}
  {\bibfnamefont {N.}~\bibnamefont {Douguet}}, \bibinfo {author} {\bibfnamefont
  {A.}~\bibnamefont {Bray}}, \bibinfo {author} {\bibfnamefont {I.}~\bibnamefont
  {Ivanov}}, \bibinfo {author} {\bibfnamefont {K.}~\bibnamefont {Bartschat}},
  \bibinfo {author} {\bibfnamefont {A.}~\bibnamefont {Kheifets}}, \bibinfo
  {author} {\bibfnamefont {R.~T.}\ \bibnamefont {Sang}},\ and\ \bibinfo
  {author} {\bibfnamefont {I.~V.}\ \bibnamefont {Litvinyuk}},\ }\bibfield
  {title} {\bibinfo {title} {Attosecond angular streaking and tunnelling time
  in atomic hydrogen},\ }\href@noop {} {\bibfield  {journal} {\bibinfo
  {journal} {Nature}\ }\textbf {\bibinfo {volume} {568}},\ \bibinfo {pages}
  {75} (\bibinfo {year} {2019})}\BibitemShut {NoStop}%
\bibitem [{\citenamefont {Ivanov}\ and\ \citenamefont
  {Kheifets}(2014)}]{Ivanov_2014}%
  \BibitemOpen
  \bibfield  {author} {\bibinfo {author} {\bibfnamefont {I.~A.}\ \bibnamefont
  {Ivanov}}\ and\ \bibinfo {author} {\bibfnamefont {A.~S.}\ \bibnamefont
  {Kheifets}},\ }\bibfield  {title} {\bibinfo {title} {{Strong-field ionization
  of He by elliptically polarized light in attoclock configuration}},\
  }\href@noop {} {\bibfield  {journal} {\bibinfo  {journal} {Phys. Rev. A}\
  }\textbf {\bibinfo {volume} {89}},\ \bibinfo {pages} {021402} (\bibinfo
  {year} {2014})}\BibitemShut {NoStop}%
\bibitem [{\citenamefont {Quan}\ \emph {et~al.}(2019)\citenamefont {Quan},
  \citenamefont {Serov}, \citenamefont {Wei}, \citenamefont {Zhao},
  \citenamefont {Zhou}, \citenamefont {Wang}, \citenamefont {Lai},
  \citenamefont {Kheifets},\ and\ \citenamefont {Liu}}]{Quan_2019}%
  \BibitemOpen
  \bibfield  {author} {\bibinfo {author} {\bibfnamefont {W.}~\bibnamefont
  {Quan}}, \bibinfo {author} {\bibfnamefont {V.~V.}\ \bibnamefont {Serov}},
  \bibinfo {author} {\bibfnamefont {M.}~\bibnamefont {Wei}}, \bibinfo {author}
  {\bibfnamefont {M.}~\bibnamefont {Zhao}}, \bibinfo {author} {\bibfnamefont
  {Y.}~\bibnamefont {Zhou}}, \bibinfo {author} {\bibfnamefont {Y.}~\bibnamefont
  {Wang}}, \bibinfo {author} {\bibfnamefont {X.}~\bibnamefont {Lai}}, \bibinfo
  {author} {\bibfnamefont {A.~S.}\ \bibnamefont {Kheifets}},\ and\ \bibinfo
  {author} {\bibfnamefont {X.}~\bibnamefont {Liu}},\ }\bibfield  {title}
  {\bibinfo {title} {Attosecond molecular angular streaking with all-ionic
  fragments detection},\ }\href@noop {} {\bibfield  {journal} {\bibinfo
  {journal} {Phys. Rev. Lett.}\ }\textbf {\bibinfo {volume} {123}},\ \bibinfo
  {pages} {223204} (\bibinfo {year} {2019})}\BibitemShut {NoStop}%
\bibitem [{\citenamefont {Serov}\ \emph {et~al.}(2019)\citenamefont {Serov},
  \citenamefont {Bray},\ and\ \citenamefont {Kheifets}}]{Serov_2019}%
  \BibitemOpen
  \bibfield  {author} {\bibinfo {author} {\bibfnamefont {V.~V.}\ \bibnamefont
  {Serov}}, \bibinfo {author} {\bibfnamefont {A.~W.}\ \bibnamefont {Bray}},\
  and\ \bibinfo {author} {\bibfnamefont {A.~S.}\ \bibnamefont {Kheifets}},\
  }\bibfield  {title} {\bibinfo {title} {Numerical attoclock on atomic and
  molecular hydrogen},\ }\href@noop {} {\bibfield  {journal} {\bibinfo
  {journal} {Phys. Rev. A}\ }\textbf {\bibinfo {volume} {99}},\ \bibinfo
  {pages} {063428} (\bibinfo {year} {2019})}\BibitemShut {NoStop}%
\bibitem [{\citenamefont {Serov}\ \emph {et~al.}(2021)\citenamefont {Serov},
  \citenamefont {Cesca},\ and\ \citenamefont {Kheifets}}]{Serov_2021}%
  \BibitemOpen
  \bibfield  {author} {\bibinfo {author} {\bibfnamefont {V.~V.}\ \bibnamefont
  {Serov}}, \bibinfo {author} {\bibfnamefont {J.}~\bibnamefont {Cesca}},\ and\
  \bibinfo {author} {\bibfnamefont {A.~S.}\ \bibnamefont {Kheifets}},\
  }\bibfield  {title} {\bibinfo {title} {Numerical and laboratory attoclock
  simulations on noble-gas atoms},\ }\href
  {https://doi.org/10.1103/PhysRevA.103.023110} {\bibfield  {journal} {\bibinfo
   {journal} {Phys. Rev. A}\ }\textbf {\bibinfo {volume} {103}},\ \bibinfo
  {pages} {023110} (\bibinfo {year} {2021})}\BibitemShut {NoStop}%
\bibitem [{\citenamefont {Goreslavski}\ \emph {et~al.}(2004)\citenamefont
  {Goreslavski}, \citenamefont {Paulus}, \citenamefont {Popruzhenko},\ and\
  \citenamefont {Shvetsov-Shilovski}}]{Goreslavski_2004}%
  \BibitemOpen
  \bibfield  {author} {\bibinfo {author} {\bibfnamefont {S.~P.}\ \bibnamefont
  {Goreslavski}}, \bibinfo {author} {\bibfnamefont {G.~G.}\ \bibnamefont
  {Paulus}}, \bibinfo {author} {\bibfnamefont {S.~V.}\ \bibnamefont
  {Popruzhenko}},\ and\ \bibinfo {author} {\bibfnamefont {N.~I.}\ \bibnamefont
  {Shvetsov-Shilovski}},\ }\bibfield  {title} {\bibinfo {title} {Coulomb
  asymmetry in above-threshold ionization},\ }\href@noop {} {\bibfield
  {journal} {\bibinfo  {journal} {Phys. Rev. Lett.}\ }\textbf {\bibinfo
  {volume} {93}},\ \bibinfo {pages} {233002} (\bibinfo {year}
  {2004})}\BibitemShut {NoStop}%
\bibitem [{\citenamefont {Boge}\ \emph {et~al.}(2013)\citenamefont {Boge},
  \citenamefont {Cirelli}, \citenamefont {Landsman}, \citenamefont {Heuser},
  \citenamefont {Ludwig}, \citenamefont {Maurer}, \citenamefont {Weger},
  \citenamefont {Gallmann},\ and\ \citenamefont {Keller}}]{Boge_2013}%
  \BibitemOpen
  \bibfield  {author} {\bibinfo {author} {\bibfnamefont {R.}~\bibnamefont
  {Boge}}, \bibinfo {author} {\bibfnamefont {C.}~\bibnamefont {Cirelli}},
  \bibinfo {author} {\bibfnamefont {A.~S.}\ \bibnamefont {Landsman}}, \bibinfo
  {author} {\bibfnamefont {S.}~\bibnamefont {Heuser}}, \bibinfo {author}
  {\bibfnamefont {A.}~\bibnamefont {Ludwig}}, \bibinfo {author} {\bibfnamefont
  {J.}~\bibnamefont {Maurer}}, \bibinfo {author} {\bibfnamefont
  {M.}~\bibnamefont {Weger}}, \bibinfo {author} {\bibfnamefont
  {L.}~\bibnamefont {Gallmann}},\ and\ \bibinfo {author} {\bibfnamefont
  {U.}~\bibnamefont {Keller}},\ }\bibfield  {title} {\bibinfo {title} {Probing
  nonadiabatic effects in strong-field tunnel ionization},\ }\href@noop {}
  {\bibfield  {journal} {\bibinfo  {journal} {Phys. Rev. Lett.}\ }\textbf
  {\bibinfo {volume} {111}},\ \bibinfo {pages} {103003} (\bibinfo {year}
  {2013})}\BibitemShut {NoStop}%
\bibitem [{\citenamefont {Bray}\ \emph {et~al.}(2018)\citenamefont {Bray},
  \citenamefont {Eckart},\ and\ \citenamefont {Kheifets}}]{Bray_2018}%
  \BibitemOpen
  \bibfield  {author} {\bibinfo {author} {\bibfnamefont {A.~W.}\ \bibnamefont
  {Bray}}, \bibinfo {author} {\bibfnamefont {S.}~\bibnamefont {Eckart}},\ and\
  \bibinfo {author} {\bibfnamefont {A.~S.}\ \bibnamefont {Kheifets}},\
  }\bibfield  {title} {\bibinfo {title} {Keldysh-rutherford model for the
  attoclock},\ }\href@noop {} {\bibfield  {journal} {\bibinfo  {journal} {Phys.
  Rev. Lett.}\ }\textbf {\bibinfo {volume} {121}},\ \bibinfo {pages} {123201}
  (\bibinfo {year} {2018})}\BibitemShut {NoStop}%
\bibitem [{\citenamefont {Douguet}\ and\ \citenamefont
  {Bartschat}(2022)}]{Douguet_2022}%
  \BibitemOpen
  \bibfield  {author} {\bibinfo {author} {\bibfnamefont {N.}~\bibnamefont
  {Douguet}}\ and\ \bibinfo {author} {\bibfnamefont {K.}~\bibnamefont
  {Bartschat}},\ }\bibfield  {title} {\bibinfo {title} {Photoelectron momentum
  distributions in the strong-field ionization of atomic hydrogen by few-cycle
  elliptically polarized optical pulses},\ }\href
  {https://doi.org/10.1103/PhysRevA.106.053112} {\bibfield  {journal} {\bibinfo
   {journal} {Phys. Rev. A}\ }\textbf {\bibinfo {volume} {106}},\ \bibinfo
  {pages} {053112} (\bibinfo {year} {2022})}\BibitemShut {NoStop}%
\bibitem [{\citenamefont {Klaiber}\ \emph {et~al.}(2015)\citenamefont
  {Klaiber}, , \citenamefont {Hatsagortsyan},\ and\ \citenamefont
  {Keitel}}]{Klaiber_2015}%
  \BibitemOpen
  \bibfield  {author} {\bibinfo {author} {\bibfnamefont {M.}~\bibnamefont
  {Klaiber}}, , \bibinfo {author} {\bibfnamefont {K.~Z.}\ \bibnamefont
  {Hatsagortsyan}},\ and\ \bibinfo {author} {\bibfnamefont {C.~H.}\
  \bibnamefont {Keitel}},\ }\bibfield  {title} {\bibinfo {title} {Tunneling
  dynamics in multiphoton ionization and attoclock calibration},\ }\href@noop
  {} {\bibfield  {journal} {\bibinfo  {journal} {Phys. Rev. Lett.}\ }\textbf
  {\bibinfo {volume} {114}},\ \bibinfo {pages} {083001} (\bibinfo {year}
  {2015})}\BibitemShut {NoStop}%
\bibitem [{\citenamefont {Xie}\ \emph {et~al.}(2024)\citenamefont {Xie},
  \citenamefont {Li}, \citenamefont {Li}, \citenamefont {Liu}, \citenamefont
  {Liu}, \citenamefont {Cao}, \citenamefont {Guo}, \citenamefont {Liu},
  \citenamefont {Zhou},\ and\ \citenamefont {Lu}}]{Xie_2024}%
  \BibitemOpen
  \bibfield  {author} {\bibinfo {author} {\bibfnamefont {W.}~\bibnamefont
  {Xie}}, \bibinfo {author} {\bibfnamefont {Z.}~\bibnamefont {Li}}, \bibinfo
  {author} {\bibfnamefont {M.}~\bibnamefont {Li}}, \bibinfo {author}
  {\bibfnamefont {Y.}~\bibnamefont {Liu}}, \bibinfo {author} {\bibfnamefont
  {Y.}~\bibnamefont {Liu}}, \bibinfo {author} {\bibfnamefont {C.}~\bibnamefont
  {Cao}}, \bibinfo {author} {\bibfnamefont {K.}~\bibnamefont {Guo}}, \bibinfo
  {author} {\bibfnamefont {K.}~\bibnamefont {Liu}}, \bibinfo {author}
  {\bibfnamefont {Y.}~\bibnamefont {Zhou}},\ and\ \bibinfo {author}
  {\bibfnamefont {P.}~\bibnamefont {Lu}},\ }\bibfield  {title} {\bibinfo
  {title} {Observation of attosecond time delays in above-threshold
  ionization},\ }\href@noop {} {\bibfield  {journal} {\bibinfo  {journal}
  {Phys. Rev. Lett.}\ }\textbf {\bibinfo {volume} {133}},\ \bibinfo {pages}
  {183201} (\bibinfo {year} {2024})}\BibitemShut {NoStop}%
\bibitem [{\citenamefont {Fu}\ \emph {et~al.}(2012)\citenamefont {Fu},
  \citenamefont {Xin}, \citenamefont {Ye},\ and\ \citenamefont
  {Liu}}]{Fu_2012}%
  \BibitemOpen
  \bibfield  {author} {\bibinfo {author} {\bibfnamefont {L.~B.}\ \bibnamefont
  {Fu}}, \bibinfo {author} {\bibfnamefont {G.~G.}\ \bibnamefont {Xin}},
  \bibinfo {author} {\bibfnamefont {D.~F.}\ \bibnamefont {Ye}},\ and\ \bibinfo
  {author} {\bibfnamefont {J.}~\bibnamefont {Liu}},\ }\bibfield  {title}
  {\bibinfo {title} {Recollision dynamics and phase diagram for nonsequential
  double ionization with circularly polarized laser fields},\ }\href
  {https://doi.org/10.1103/PhysRevLett.108.103601} {\bibfield  {journal}
  {\bibinfo  {journal} {Phys. Rev. Lett.}\ }\textbf {\bibinfo {volume} {108}},\
  \bibinfo {pages} {103601} (\bibinfo {year} {2012})}\BibitemShut {NoStop}%
\bibitem [{\citenamefont {Pfeiffer}\ \emph
  {et~al.}(2012{\natexlab{b}})\citenamefont {Pfeiffer}, \citenamefont
  {Cirelli}, \citenamefont {Landsman}, \citenamefont {Smolarski}, \citenamefont
  {Dimitrovski}, \citenamefont {Madsen},\ and\ \citenamefont
  {Keller}}]{Pfeiffer_2012prl}%
  \BibitemOpen
  \bibfield  {author} {\bibinfo {author} {\bibfnamefont {A.~N.}\ \bibnamefont
  {Pfeiffer}}, \bibinfo {author} {\bibfnamefont {C.}~\bibnamefont {Cirelli}},
  \bibinfo {author} {\bibfnamefont {A.~S.}\ \bibnamefont {Landsman}}, \bibinfo
  {author} {\bibfnamefont {M.}~\bibnamefont {Smolarski}}, \bibinfo {author}
  {\bibfnamefont {D.}~\bibnamefont {Dimitrovski}}, \bibinfo {author}
  {\bibfnamefont {L.~B.}\ \bibnamefont {Madsen}},\ and\ \bibinfo {author}
  {\bibfnamefont {U.}~\bibnamefont {Keller}},\ }\bibfield  {title} {\bibinfo
  {title} {Probing the longitudinal momentum spread of the electron wave packet
  at the tunnel exit},\ }\href {https://doi.org/10.1103/PhysRevLett.109.083002}
  {\bibfield  {journal} {\bibinfo  {journal} {Phys. Rev. Lett.}\ }\textbf
  {\bibinfo {volume} {109}},\ \bibinfo {pages} {083002} (\bibinfo {year}
  {2012}{\natexlab{b}})}\BibitemShut {NoStop}%
\bibitem [{\citenamefont {Li}\ \emph {et~al.}(2016)\citenamefont {Li},
  \citenamefont {Geng}, \citenamefont {Han}, \citenamefont {Liu}, \citenamefont
  {Peng}, \citenamefont {Gong},\ and\ \citenamefont {Liu}}]{Li_2016}%
  \BibitemOpen
  \bibfield  {author} {\bibinfo {author} {\bibfnamefont {M.}~\bibnamefont
  {Li}}, \bibinfo {author} {\bibfnamefont {J.-W.}\ \bibnamefont {Geng}},
  \bibinfo {author} {\bibfnamefont {M.}~\bibnamefont {Han}}, \bibinfo {author}
  {\bibfnamefont {M.-M.}\ \bibnamefont {Liu}}, \bibinfo {author} {\bibfnamefont
  {L.-Y.}\ \bibnamefont {Peng}}, \bibinfo {author} {\bibfnamefont
  {Q.}~\bibnamefont {Gong}},\ and\ \bibinfo {author} {\bibfnamefont
  {Y.}~\bibnamefont {Liu}},\ }\bibfield  {title} {\bibinfo {title} {Subcycle
  nonadiabatic strong-field tunneling ionization},\ }\href
  {https://doi.org/10.1103/PhysRevA.93.013402} {\bibfield  {journal} {\bibinfo
  {journal} {Phys. Rev. A}\ }\textbf {\bibinfo {volume} {93}},\ \bibinfo
  {pages} {013402} (\bibinfo {year} {2016})}\BibitemShut {NoStop}%
\bibitem [{\citenamefont {Han}\ \emph {et~al.}(2019)\citenamefont {Han},
  \citenamefont {Ge}, \citenamefont {Fang}, \citenamefont {Yu}, \citenamefont
  {Guo}, \citenamefont {Ma}, \citenamefont {Deng}, \citenamefont {Gong},\ and\
  \citenamefont {Liu}}]{Han_2019}%
  \BibitemOpen
  \bibfield  {author} {\bibinfo {author} {\bibfnamefont {M.}~\bibnamefont
  {Han}}, \bibinfo {author} {\bibfnamefont {P.}~\bibnamefont {Ge}}, \bibinfo
  {author} {\bibfnamefont {Y.}~\bibnamefont {Fang}}, \bibinfo {author}
  {\bibfnamefont {X.}~\bibnamefont {Yu}}, \bibinfo {author} {\bibfnamefont
  {Z.}~\bibnamefont {Guo}}, \bibinfo {author} {\bibfnamefont {X.}~\bibnamefont
  {Ma}}, \bibinfo {author} {\bibfnamefont {Y.}~\bibnamefont {Deng}}, \bibinfo
  {author} {\bibfnamefont {Q.}~\bibnamefont {Gong}},\ and\ \bibinfo {author}
  {\bibfnamefont {Y.}~\bibnamefont {Liu}},\ }\bibfield  {title} {\bibinfo
  {title} {Unifying tunneling pictures of strong-field ionization with an
  improved attoclock},\ }\href@noop {} {\bibfield  {journal} {\bibinfo
  {journal} {Phys. Rev. Lett.}\ }\textbf {\bibinfo {volume} {123}},\ \bibinfo
  {pages} {073201} (\bibinfo {year} {2019})}\BibitemShut {NoStop}%
\bibitem [{\citenamefont {Dubois}\ \emph {et~al.}(2020)\citenamefont {Dubois},
  \citenamefont {Chandre},\ and\ \citenamefont {Uzer}}]{Dubois_2020a}%
  \BibitemOpen
  \bibfield  {author} {\bibinfo {author} {\bibfnamefont {J.}~\bibnamefont
  {Dubois}}, \bibinfo {author} {\bibfnamefont {C.}~\bibnamefont {Chandre}},\
  and\ \bibinfo {author} {\bibfnamefont {T.}~\bibnamefont {Uzer}},\ }\bibfield
  {title} {\bibinfo {title} {Nonadiabatic effects in the double ionization of
  atoms driven by a circularly polarized laser pulse},\ }\href
  {https://doi.org/10.1103/PhysRevE.102.032218} {\bibfield  {journal} {\bibinfo
   {journal} {Phys. Rev. E}\ }\textbf {\bibinfo {volume} {102}},\ \bibinfo
  {pages} {032218} (\bibinfo {year} {2020})}\BibitemShut {NoStop}%
\bibitem [{\citenamefont {Trabert}\ \emph {et~al.}(2021)\citenamefont
  {Trabert}, \citenamefont {Anders}, \citenamefont {Brennecke}, \citenamefont
  {Sch\"offler}, \citenamefont {Jahnke}, \citenamefont {Schmidt}, \citenamefont
  {Kunitski}, \citenamefont {Lein}, \citenamefont {D\"orner},\ and\
  \citenamefont {Eckart}}]{Trabert_2021}%
  \BibitemOpen
  \bibfield  {author} {\bibinfo {author} {\bibfnamefont {D.}~\bibnamefont
  {Trabert}}, \bibinfo {author} {\bibfnamefont {N.}~\bibnamefont {Anders}},
  \bibinfo {author} {\bibfnamefont {S.}~\bibnamefont {Brennecke}}, \bibinfo
  {author} {\bibfnamefont {M.~S.}\ \bibnamefont {Sch\"offler}}, \bibinfo
  {author} {\bibfnamefont {T.}~\bibnamefont {Jahnke}}, \bibinfo {author}
  {\bibfnamefont {L.~P.~H.}\ \bibnamefont {Schmidt}}, \bibinfo {author}
  {\bibfnamefont {M.}~\bibnamefont {Kunitski}}, \bibinfo {author}
  {\bibfnamefont {M.}~\bibnamefont {Lein}}, \bibinfo {author} {\bibfnamefont
  {R.}~\bibnamefont {D\"orner}},\ and\ \bibinfo {author} {\bibfnamefont
  {S.}~\bibnamefont {Eckart}},\ }\bibfield  {title} {\bibinfo {title}
  {Nonadiabatic strong field ionization of atomic hydrogen},\ }\href
  {https://doi.org/10.1103/PhysRevLett.127.273201} {\bibfield  {journal}
  {\bibinfo  {journal} {Phys. Rev. Lett.}\ }\textbf {\bibinfo {volume} {127}},\
  \bibinfo {pages} {273201} (\bibinfo {year} {2021})}\BibitemShut {NoStop}%
\bibitem [{\citenamefont {Mauger}\ \emph
  {et~al.}(2010{\natexlab{a}})\citenamefont {Mauger}, \citenamefont {Chandre},\
  and\ \citenamefont {Uzer}}]{Mauger_2010a}%
  \BibitemOpen
  \bibfield  {author} {\bibinfo {author} {\bibfnamefont {F.}~\bibnamefont
  {Mauger}}, \bibinfo {author} {\bibfnamefont {C.}~\bibnamefont {Chandre}},\
  and\ \bibinfo {author} {\bibfnamefont {T.}~\bibnamefont {Uzer}},\ }\bibfield
  {title} {\bibinfo {title} {From recollisions to the knee: A road map for
  double ionization in intense laser fields},\ }\href@noop {} {\bibfield
  {journal} {\bibinfo  {journal} {Phys. Rev. Lett.}\ }\textbf {\bibinfo
  {volume} {104}},\ \bibinfo {pages} {043005} (\bibinfo {year}
  {2010}{\natexlab{a}})}\BibitemShut {NoStop}%
\bibitem [{\citenamefont {Mauger}\ \emph
  {et~al.}(2010{\natexlab{b}})\citenamefont {Mauger}, \citenamefont {Chandre},\
  and\ \citenamefont {Uzer}}]{Mauger_2010b}%
  \BibitemOpen
  \bibfield  {author} {\bibinfo {author} {\bibfnamefont {F.}~\bibnamefont
  {Mauger}}, \bibinfo {author} {\bibfnamefont {C.}~\bibnamefont {Chandre}},\
  and\ \bibinfo {author} {\bibfnamefont {T.}~\bibnamefont {Uzer}},\ }\bibfield
  {title} {\bibinfo {title} {Recollisions and correlated double ionization with
  circularly polarized light},\ }\href@noop {} {\bibfield  {journal} {\bibinfo
  {journal} {Phys. Rev. Lett.}\ }\textbf {\bibinfo {volume} {105}},\ \bibinfo
  {pages} {083002} (\bibinfo {year} {2010}{\natexlab{b}})}\BibitemShut
  {NoStop}%
\bibitem [{\citenamefont {Kamor}\ \emph {et~al.}(2013)\citenamefont {Kamor},
  \citenamefont {Mauger}, \citenamefont {Chandre},\ and\ \citenamefont
  {Uzer}}]{Mauger_2013}%
  \BibitemOpen
  \bibfield  {author} {\bibinfo {author} {\bibfnamefont {A.}~\bibnamefont
  {Kamor}}, \bibinfo {author} {\bibfnamefont {F.}~\bibnamefont {Mauger}},
  \bibinfo {author} {\bibfnamefont {C.}~\bibnamefont {Chandre}},\ and\ \bibinfo
  {author} {\bibfnamefont {T.}~\bibnamefont {Uzer}},\ }\bibfield  {title}
  {\bibinfo {title} {How key periodic orbits drive recollisions in a circularly
  polarized laser field},\ }\href@noop {} {\bibfield  {journal} {\bibinfo
  {journal} {Phys. Rev. Lett.}\ }\textbf {\bibinfo {volume} {110}},\ \bibinfo
  {pages} {253002} (\bibinfo {year} {2013})}\BibitemShut {NoStop}%
\bibitem [{\citenamefont {Heldt}\ \emph {et~al.}(2023)\citenamefont {Heldt},
  \citenamefont {Dubois}, \citenamefont {Birk}, \citenamefont {Borisova},
  \citenamefont {Lando}, \citenamefont {Ott},\ and\ \citenamefont
  {Pfeifer}}]{Heldt_2023}%
  \BibitemOpen
  \bibfield  {author} {\bibinfo {author} {\bibfnamefont {T.}~\bibnamefont
  {Heldt}}, \bibinfo {author} {\bibfnamefont {J.}~\bibnamefont {Dubois}},
  \bibinfo {author} {\bibfnamefont {P.}~\bibnamefont {Birk}}, \bibinfo {author}
  {\bibfnamefont {G.~D.}\ \bibnamefont {Borisova}}, \bibinfo {author}
  {\bibfnamefont {G.~M.}\ \bibnamefont {Lando}}, \bibinfo {author}
  {\bibfnamefont {C.}~\bibnamefont {Ott}},\ and\ \bibinfo {author}
  {\bibfnamefont {T.}~\bibnamefont {Pfeifer}},\ }\bibfield  {title} {\bibinfo
  {title} {Attosecond real-time observation of recolliding electron
  trajectories in helium at low laser intensities},\ }\href
  {https://doi.org/10.1103/PhysRevLett.130.183201} {\bibfield  {journal}
  {\bibinfo  {journal} {Phys. Rev. Lett.}\ }\textbf {\bibinfo {volume} {130}},\
  \bibinfo {pages} {183201} (\bibinfo {year} {2023})}\BibitemShut {NoStop}%
\bibitem [{SM()}]{SM}%
  \BibitemOpen
  \href@noop {} {}\bibinfo {howpublished} {See the Supplemental Materials for
  the details of the applied techniques.}\BibitemShut {Stop}%
\bibitem [{\citenamefont {Klaiber}\ \emph
  {et~al.}(2022{\natexlab{a}})\citenamefont {Klaiber}, \citenamefont {Lv},
  \citenamefont {Hatsagortsyan},\ and\ \citenamefont
  {Keitel}}]{Klaiber_2022edge}%
  \BibitemOpen
  \bibfield  {author} {\bibinfo {author} {\bibfnamefont {M.}~\bibnamefont
  {Klaiber}}, \bibinfo {author} {\bibfnamefont {Q.~Z.}\ \bibnamefont {Lv}},
  \bibinfo {author} {\bibfnamefont {K.~Z.}\ \bibnamefont {Hatsagortsyan}},\
  and\ \bibinfo {author} {\bibfnamefont {C.~H.}\ \bibnamefont {Keitel}},\
  }\bibfield  {title} {\bibinfo {title} {Tunneling ionization in ultrashort
  laser pulses: Edge effect and remedy},\ }\href
  {https://doi.org/10.1103/PhysRevA.105.063109} {\bibfield  {journal} {\bibinfo
   {journal} {Phys. Rev. A}\ }\textbf {\bibinfo {volume} {105}},\ \bibinfo
  {pages} {063109} (\bibinfo {year} {2022}{\natexlab{a}})}\BibitemShut
  {NoStop}%
\bibitem [{\citenamefont {Popov}(2004)}]{Popov_2004u}%
  \BibitemOpen
  \bibfield  {author} {\bibinfo {author} {\bibfnamefont {V.~S.}\ \bibnamefont
  {Popov}},\ }\bibfield  {title} {\bibinfo {title} {{Tunneling and multiphoton
  ionization of atoms and ions ina strong laser field (Keldysh theory)}},\
  }\href@noop {} {\bibfield  {journal} {\bibinfo  {journal} {Phys. Usp.}\
  }\textbf {\bibinfo {volume} {47}},\ \bibinfo {pages} {855} (\bibinfo {year}
  {2004})}\BibitemShut {NoStop}%
\bibitem [{\citenamefont {Mur}\ \emph {et~al.}(2001)\citenamefont {Mur},
  \citenamefont {Popruzhenko},\ and\ \citenamefont {Popov}}]{Mur_2001}%
  \BibitemOpen
  \bibfield  {author} {\bibinfo {author} {\bibfnamefont {V.~D.}\ \bibnamefont
  {Mur}}, \bibinfo {author} {\bibfnamefont {S.~V.}\ \bibnamefont
  {Popruzhenko}},\ and\ \bibinfo {author} {\bibfnamefont {V.~S.}\ \bibnamefont
  {Popov}},\ }\bibfield  {title} {\bibinfo {title} {Energy and pulse spectra of
  photoelectrons during ionization by strong laser radiation (elliptical
  polarization)},\ }\href@noop {} {\bibfield  {journal} {\bibinfo  {journal}
  {Zh. Exp. Theor. Fiz.}\ }\textbf {\bibinfo {volume} {119}},\ \bibinfo {pages}
  {893} (\bibinfo {year} {2001})}\BibitemShut {NoStop}%
\bibitem [{\citenamefont {Klaiber}\ \emph {et~al.}(2013)\citenamefont
  {Klaiber}, \citenamefont {Yakaboylu},\ and\ \citenamefont
  {Hatsagortsyan}}]{Klaiber_2013a}%
  \BibitemOpen
  \bibfield  {author} {\bibinfo {author} {\bibfnamefont {M.}~\bibnamefont
  {Klaiber}}, \bibinfo {author} {\bibfnamefont {E.}~\bibnamefont {Yakaboylu}},\
  and\ \bibinfo {author} {\bibfnamefont {K.~Z.}\ \bibnamefont
  {Hatsagortsyan}},\ }\bibfield  {title} {\bibinfo {title} {{Above-threshold
  ionization with highly charged ions in superstrong laser fields. I.
  Coulomb-corrected strong-field approximation}},\ }\href@noop {} {\bibfield
  {journal} {\bibinfo  {journal} {Phys. Rev. A}\ }\textbf {\bibinfo {volume}
  {87}},\ \bibinfo {pages} {023417} (\bibinfo {year} {2013})}\BibitemShut
  {NoStop}%
\bibitem [{\citenamefont {Klaiber}\ \emph
  {et~al.}(2022{\natexlab{b}})\citenamefont {Klaiber}, \citenamefont {Lv},
  \citenamefont {Sukiasyan}, \citenamefont {Bakucz~Can\'ario}, \citenamefont
  {Hatsagortsyan},\ and\ \citenamefont {Keitel}}]{Klaiber_2022R}%
  \BibitemOpen
  \bibfield  {author} {\bibinfo {author} {\bibfnamefont {M.}~\bibnamefont
  {Klaiber}}, \bibinfo {author} {\bibfnamefont {Q.~Z.}\ \bibnamefont {Lv}},
  \bibinfo {author} {\bibfnamefont {S.}~\bibnamefont {Sukiasyan}}, \bibinfo
  {author} {\bibfnamefont {D.}~\bibnamefont {Bakucz~Can\'ario}}, \bibinfo
  {author} {\bibfnamefont {K.~Z.}\ \bibnamefont {Hatsagortsyan}},\ and\
  \bibinfo {author} {\bibfnamefont {C.~H.}\ \bibnamefont {Keitel}},\ }\bibfield
   {title} {\bibinfo {title} {Reconciling conflicting approaches for the
  tunneling time delay in strong field ionization},\ }\href
  {https://doi.org/10.1103/PhysRevLett.129.203201} {\bibfield  {journal}
  {\bibinfo  {journal} {Phys. Rev. Lett.}\ }\textbf {\bibinfo {volume} {129}},\
  \bibinfo {pages} {203201} (\bibinfo {year} {2022}{\natexlab{b}})}\BibitemShut
  {NoStop}%
\bibitem [{\citenamefont {Yu}\ \emph {et~al.}(2022)\citenamefont {Yu},
  \citenamefont {Liu}, \citenamefont {Li}, \citenamefont {Yan}, \citenamefont
  {Cao}, \citenamefont {Tan}, \citenamefont {Liang}, \citenamefont {Guo},
  \citenamefont {Cao}, \citenamefont {Lan}, \citenamefont {Zhang},
  \citenamefont {Zhou},\ and\ \citenamefont {Lu}}]{Yu_2022}%
  \BibitemOpen
  \bibfield  {author} {\bibinfo {author} {\bibfnamefont {M.}~\bibnamefont
  {Yu}}, \bibinfo {author} {\bibfnamefont {K.}~\bibnamefont {Liu}}, \bibinfo
  {author} {\bibfnamefont {M.}~\bibnamefont {Li}}, \bibinfo {author}
  {\bibfnamefont {J.}~\bibnamefont {Yan}}, \bibinfo {author} {\bibfnamefont
  {C.}~\bibnamefont {Cao}}, \bibinfo {author} {\bibfnamefont {J.}~\bibnamefont
  {Tan}}, \bibinfo {author} {\bibfnamefont {J.}~\bibnamefont {Liang}}, \bibinfo
  {author} {\bibfnamefont {K.}~\bibnamefont {Guo}}, \bibinfo {author}
  {\bibfnamefont {W.}~\bibnamefont {Cao}}, \bibinfo {author} {\bibfnamefont
  {P.}~\bibnamefont {Lan}}, \bibinfo {author} {\bibfnamefont {Q.}~\bibnamefont
  {Zhang}}, \bibinfo {author} {\bibfnamefont {Y.}~\bibnamefont {Zhou}},\ and\
  \bibinfo {author} {\bibfnamefont {P.}~\bibnamefont {Lu}},\ }\bibfield
  {title} {\bibinfo {title} {Full experimental determination of tunneling time
  with attosecond-scale streaking method},\ }\href@noop {} {\bibfield
  {journal} {\bibinfo  {journal} {Light: Science \& Applications}\ }\textbf
  {\bibinfo {volume} {11}},\ \bibinfo {pages} {215} (\bibinfo {year}
  {2022})}\BibitemShut {NoStop}%
\end{thebibliography}%

\end{document}